\documentclass[a4paper,UKenglish]{lipics}
 
\usepackage{microtype}

\title{Rank logic is dead, long live rank logic!}

\author{Erich Grädel}
\author{Wied Pakusa}
\affil{Mathematical Foundations of Computer Science, RWTH Aachen University\\
  \texttt{\{graedel,pakusa\}@logic.rwth-aachen.de}}
\authorrunning{E.\,Grädel and W.\,Pakusa}
\Copyright{Erich Grädel and Wied Pakusa}

\usepackage{mathtools}
\usepackage{fixltx2e}
\usepackage{xspace}
\usepackage{xcolor}
\usepackage{enumerate}
\usepackage{amsmath}
\usepackage{amsfonts}
\usepackage{MnSymbol}
\usepackage{amsthm}
\usepackage{booktabs}
\usepackage{blindtext}
\usepackage{bbm}
\usepackage{graphicx}
\usepackage{todonotes}

\usepackage[]{fixme}

\usepackage{tikz}
\usetikzlibrary{arrows,shapes,positioning,arrows,decorations,calc,
decorations.pathmorphing,decorations.pathreplacing}
\tikzset{every picture/.style={>=stealth,thick}}

\makeatletter
\newcommand{\@abbrev}[3]{
  \def\c@a@def##1{
      \if ##1.
        \relax
      \else
        \@ifdefinable{\@nameuse{#1##1}}{\@namedef{#1##1}{#2##1}}
        \expandafter\c@a@def
      \fi
    }
  \c@a@def #3.
}
\@abbrev{bb}{\mathbb}{ABCDEFGHIJKLMNOPQRSTUVWXYZ}
\@abbrev{bf}{\mathbf}{ABCDEFGHIJKLMNOPQRSTUVWXYZabcdefghijklmnopqrstuvwxyz}
\@abbrev{bit}{\boldsymbol}{ABCDEFGHIJKLMNOPQRSTUVWXYZabcdefghijklmnopqrstuvwxyz}
\@abbrev{mc}{\mathcal}{ABCDEFGHIJKLMNOPQRSTUVWXYZ}
\@abbrev{mf}{\mathfrak}{ABCDEFGHIJKLMNOPQRSTUVWXYZabcdefghijklmnopqrstuvwxyz}
\@abbrev{rm}{\mathrm}{ABCDEFGHIJKLMNOPQRSTUVWXYZabcdefghijklmnopqrstuvwxyz}
\@abbrev{scr}{\mathscr}{ABCDEFGHIJKLMNOPQRSTUVWXYZabcdefghijklmnopqrstuvwxyz}
\@abbrev{sf}{\mathsf}{ABCDEFGHIJKLMNOPQRSTUVWXYZabcdefghijklmnopqrstuvwxyz}
\makeatother

\newcommand{\defeq}{:=}

\newcommand{\isom}{\cong}
\newcommand{\nisom}{\not\cong}

\newcommand{\inseg}[1]{\ensuremath{[#1]}}
\newcommand{\card}[1]{\ensuremath{|#1|}}
\newcommand{\sgn}{\ensuremath{\textrm{sgn}}}
\newcommand{\modulo}{\ensuremath{\text{mod }}}
\newcommand{\dom}{\ensuremath{\textrm{dom}}}
\newcommand{\Primes}{\bbP}

\newcommand{\gengroup}[1]{\langle #1 \rangle}

\newcommand{\pcycle}[1]{\ensuremath{( \, #1 \, )}\xspace}

\newcommand{\field}[1]{\mathbb{#1}}

\newcommand{\Aut}{\ensuremath{\text{Aut}}\xspace}
\newcommand{\Sym}{\ensuremath{\text{Sym}}\xspace}
\newcommand{\Iso}{\ensuremath{\text{Iso}}\xspace}

\newcommand{\con}{\ensuremath{\text{con}}\xspace}

\newcommand{\onevec}{\ensuremath{\mathbbm{1}}}
\newcommand{\zerovec}{\ensuremath{\vec{0}}}

\newcommand{\N}{\ensuremath{\bbN}}
\newcommand{\Z}{\ensuremath{\bbZ}}

\newcommand{\Str}{\ensuremath{\mathscr{S}}}

\newcommand{\logic}[1]{\ensuremath{\textsc{#1}}\xspace}

\newcommand{\FO}{\logic{FO}}
\newcommand{\FOC}{\logic{FOC}}

\newcommand{\STC}{\logic{STC}}

\newcommand{\FP}{\logic{FP}}

\newcommand{\FPC}{\logic{FPC}}
\newcommand{\FPRK}{\logic{FPR}}
\newcommand{\FPR}{\FPRK}

\newcommand{\FOSp}{\ensuremath{\logic{FOS}_p}\xspace}
\newcommand{\FOSx}[1]{\ensuremath{\logic{FOS}_{#1}}\xspace}
\newcommand{\FPSp}{\ensuremath{\logic{FPS}_p}\xspace}
\newcommand{\FPS}{\ensuremath{\logic{FPS}}\xspace}
\newcommand{\FPSvar}{\ensuremath{\logic{FPS}^*}\xspace}
\newcommand{\FPSx}[1]{\ensuremath{\logic{FPS}_{#1}}\xspace}
\newcommand{\FORp}{\ensuremath{\logic{FOR}_p}\xspace}
\newcommand{\FOR}{\ensuremath{\logic{FOR}}\xspace}
\newcommand{\FORx}[1]{\ensuremath{\logic{FOR}_{#1}}\xspace}
\newcommand{\FPRp}{\ensuremath{\logic{FPR}_p}\xspace}
\newcommand{\FPRvar}{\ensuremath{\logic{FPR}^*}\xspace}
\newcommand{\FPRx}[1]{\ensuremath{\logic{FPR}_{#1}}\xspace}

\newcommand{\INFCk}{\ensuremath{\logic{C}^k_{\infty\omega}}\xspace}
\newcommand{\INFCkx}[1]{\ensuremath{\logic{C}^{#1}_{\infty\omega}}\xspace}

\newcommand{\ifp}{\ensuremath{\textsf{ifp}}\xspace}
\newcommand{\slv}{\ensuremath{\textsf{slv}}\xspace}
\newcommand{\slvp}{\ensuremath{\textsf{slv}_p}\xspace}
\newcommand{\rkp}{\ensuremath{\textsf{rk}_p}\xspace}
\newcommand{\rkx}[1]{\ensuremath{\textsf{rk}_{#1}}\xspace}
\newcommand{\rk}{\ensuremath{\textsf{rk}^*}\xspace}
\newcommand{\rankk}{\ensuremath{\textsf{rk}}\xspace}

\newcommand{\Ckeqv}{\ensuremath{\equiv^C_k}\xspace}
\newcommand{\Ckeqvx}[1]{\ensuremath{\equiv^C_{#1}}\xspace}

\newcommand{\compclass}[1]{\ensuremath{\textsc{#1}}\xspace}
\newcommand{\PTIME}{\compclass{Ptime}}

\newcommand{\MODLx}[1]{\ensuremath{\textsc{MOD}_{#1}\textsc{L}}\xspace}

\renewcommand{\phi}{\varphi}

\newcommand{\ba}{{\bar a}}
\newcommand{\bb}{{\bar b}}
\newcommand{\bc}{{\bar c}}
\newcommand{\bd}{{\bar d}}

\newcommand{\bm}{{\bar m}}
\newcommand{\bn}{{\bar n}}
\newcommand{\bp}{{\bar p}}
\newcommand{\bv}{{\bar v}}
\newcommand{\bx}{{\bar x}}
\newcommand{\by}{{\bar y}}
\newcommand{\bz}{{\bar z}}
\newcommand{\bmu}{{\bar \mu}}
\newcommand{\bnu}{{\bar \nu}}

\newcommand{\vct}[1]{\vec{#1}}	

\newcommand{\complexityclass}[1]{\ensuremath{\textsc{#1}}\xspace}
\newcommand{\LOGSPACE}{\complexityclass{Logspace}}

\colorlet{blau}{blue!50}
\newcommand{\lightercolor}[3]{
    \colorlet{#3}{#1!#2!white}
}

\lightercolor{blau}{50}{hellblau}

\theoremstyle{plain}
\newtheorem{prop}[theorem]{Proposition}

\newcommand{\eqcount}{\ensuremath{
\stepcounter{equation}
\tag{E \theequation}
}
}

\newcommand{\eqcountreset}
{
\setcounter{equation}{0}
}
\newenvironment{claim}[1]{\par\noindent\textbf{Claim:}\space 
{#1}}{}
\newenvironment{claimproof}[1]{\par\noindent\textit{Proof of 
claim:}\space#1}
{\leavevmode\unskip\penalty9999 \hbox{}\nobreak\hfill\quad\hbox{
$\dashv$}}

\begin{document}

\maketitle

\begin{abstract}
Motivated by the search for a logic for polynomial time, we study 
rank logic (\FPR) which extends fixed-point logic with counting 
(\FPC) by operators that determine the rank of matrices over finite fields.
While  $\FPR$ can express most of the known queries that 
separate $\FPC$ from $\PTIME$, nearly nothing was known about the limitations 
of its expressive power.

In our first main result we show that the extensions of \FPC by rank operators 
over different prime fields are incomparable. This 
solves an open question posed by Dawar and Holm and also implies that rank 
logic, in its original definition with a distinct rank operator for every 
field, 
fails to capture polynomial time.
In particular we show that the variant of rank logic $\FPRvar$ with an operator 
that uniformly expresses the matrix rank over finite fields is more expressive 
than $\FPR$.

One important step in our proof is to consider solvability logic \FPS which is 
the analogous extension of \FPC by quantifiers which express the solvability 
problem for linear equation systems over finite fields.
Solvability logic can easily be embedded into rank logic, but it is open 
whether it is a strict fragment.
In our second main result we give a partial answer to this question: 
in the absence of  counting, rank operators are strictly more expressive than 
solvability quantifiers.
\end{abstract}

\section{Introduction}

\emph{``Le roi est mort, vive le roi!''} has been the traditional proclamation,
in France and other countries,
to announce not only the death of the monarch, but also the
immediate installment of his successor on the throne.
The purpose of this paper is to kill the rank logic \FPR, in the form in which 
it
has been proposed in \cite{DaGrHoLa09},  as a candidate for a logic for \PTIME.
The logic \FPR extends fixed-point logic by operators $\rkp$ (for every
prime $p$) which compute the rank of definable matrices over the prime
field $\field F_p$ with $p$ elements. Although rank logic is well-motivated,
as a logic that strictly extends fixed-point logic with counting
by the ability to express important properties of linear algebra,
most notably the solvability of linear equation systems over finite
fields, our results show that the choice of having a separate rank operator 
for every prime $p$ leads to a significant deficiency of the logic.
Indeed, it follows from our main theorem that 
even the uniform rank problem, of computing the rank of a 
given matrix over an arbitrary prime, cannot be expressed in \FPR
and thus separates $\FPR$ from \PTIME. 
This also reveals that a more general variant of rank logic,
which has already been proposed in \cite{Ho10, La11, Pa10}
and which is based on a rank operator that takes not only the matrix but also
the prime $p$ as part of the input, is indeed strictly
more powerful than $\FPR$. 
Our result thus installs this
new rank logic, denoted \FPRvar, as the rightful 
and distinctly more powerful successor of 
\FPR as a potential candidate for a logic for \PTIME. 

\subparagraph*{A logic for polynomial time}
The question whether there is a logic that expresses precisely the 
polynomial-time properties of finite structures
is an important challenge in the field of finite model theory 
\cite{FMTbook,Gr08}. 
The logic of reference for this quest is
fixed-point logic with counting (\FPC) which
captures polynomial time on many interesting classes
of structures and is strong enough to express most of 
the algorithmic techniques leading to polynomial-time procedures~\cite{Da15}.
Although it has been known for more than twenty years that
\FPC fails to capture \PTIME in general, by the fundamental
CFI-construction due to Cai, Fürer, and Immerman~\cite{CFI92},
we still do not know many properties of finite structures
that provably separate \FPC from \PTIME.
The two main sources of such problems
are tractable cases of the graph isomorphism problem and queries from the field
of linear algebra.
First of all, the CFI-construction shows that \FPC 
cannot define the isomorphism problem on graphs with bounded 
degree \emph{and}  bounded colour class size
whereas the isomorphism problem is known to
be tractable on all classes of graphs  
with bounded degree \emph{or} bounded colour class size. 
Secondly, Atserias, Bulatov and Dawar~\cite{AtBuDa09} proved 
that \FPC cannot express the solvability of linear equation systems over 
any finite Abelian group. 
It follows, that also other problems from the field of linear 
algebra are not definable in \FPC.
Interestingly, also the CFI-query can be formulated
as linear equation system over $\field F_2$~\cite{DaGrHoLa09}.

\subparagraph*{Rank logic}
This latter observation motivated Dawar, Grohe, Holm and 
Laubner~\cite{DaGrHoLa09} 
to introduce \emph{rank logic} (\FPR) which is the extension of \FPC by 
operators for the rank of definable matrices over prime 
fields $\field F_p$.
To illustrate the idea of rank logic, let
$\phi(x,y)$ be a formula (of $\FPC$, say) which defines a binary 
relation $\phi^\mfA \subseteq A \times A$ in an input structure $\mfA$.
We identify the relation $\phi^\mfA$ with the associated adjacency matrix 
\[ 
M_\phi^\mfA: A \times A \to \{ 0, 1\}, (a,b) \mapsto 
\begin{cases}
 1,& \text{ if } (a,b) \in \phi^\mfA\\
 0, & \text{ if } (a,b) \nin \phi^\mfA.
\end{cases}
\]
In this sense, the formula $\phi$ defines in every structure $\mfA$ a 
matrix $M_\phi^\mfA$ with entries in $\{0,1 \} \subseteq \field F_p$. 
Now, rank logic $\FPR$ contains for every prime 
$p \in \bbP$ a \emph{rank operator} $\rkp$ which can be used to form a 
\emph{rank term} $[ \rkp \,\,\phi(x,y) ]$ whose value in an input structure 
$\mfA$ is 
the matrix rank of $M_\phi$ over $\field F_p$ (we remark that rank logic 
also allows to express the rank of matrices which are indexed by tuples of 
elements; the precise definition is given in Section~\ref{sec:logics:linalgop}).

It turns out that rank operators have quite surprising expressive power. 
For example, they can define the transitive closure of symmetric 
relations, they can count the number of paths in DAGs modulo $p$ and they 
can express the solvability of linear equation systems over finite fields 
(recall that a linear equation system $M \cdot \vct x = \vct b$ is solvable if, 
and only if, $\rankk(M) = \rankk(M \, | \, \vct b)$)~\cite{DaGrHoLa09}. 
Furthermore, rank operators can be used to define the isomorphism problem on
various classes of structures on which the Weisfeiler-Lehman method (and thus 
fixed-point logic with counting) fails, e.g.\ classes of
C(ai)-F(ürer)-I(mmerman) graphs~\cite{CFI92,DaGrHoLa09} and 
multipedes~\cite{GuSh96,Ho10}.
The common idea of these isomorphism procedures is to reduce the isomorphism 
problem of structures to a suitable linear equation system over a finite field.
More generally, by a recent result (which is mainly concerned with another 
candidate of a logic for polynomial time~\cite{AGGP14}), it follows that \FPR 
captures polynomial time on certain classes of structures of bounded colour 
class size. In particular, this holds for the class of all structures of colour 
class size two (to which CFI-graphs and multipedes belong).

\medskip
While these results clearly show the high potential of rank 
logic, almost nothing has been known about its limitations.
For instance, it has remained open whether rank logic suffices to capture 
polynomial time, whether rank operators can simulate fixed-point 
inductions~\cite{DaGrHoLa09} and also whether rank logic 
can define closely related problems from linear algebra 
(such as the solvability of linear equations over finite 
\emph{rings} rather than fields~\cite{DaGrHoKoPa13}).
One particular intriguing question is whether rank operators over 
different prime fields can simulate each other. In other words:
 is it possible to reduce the problem of determining the rank of a matrix over 
$\field F_p$ (within fixed-point logic with counting) to the 
problem of determining the rank of a matrix over $\field F_q$ (where $p, q$ are 
distinct primes)?
To attack this problem, Dawar and Holm~\cite{DaHo12, Ho10} 
developed a powerful toolkit of so called \emph{partition games} of which one 
variant (so called \emph{matrix-equivalence games}) precisely characterises the 
expressive power of infinitary logic extended by rank quantifiers.
By using these games, Holm~\cite{Ho10} was able to give a negative answer to 
the 
above question for the restricted case of rank operators of dimension one.

\smallskip
In this paper we propose a different method, based 
on exploiting symmetries rather than game theoretic arguments, to prove new 
lower bounds for logics with rank operators.
In our main result (Theorem~\ref{thm:fps:distinct:primes}) we prove that for 
every prime $q$ there exists a class of structures $\mcK_q$ on which \FPC fails 
to capture polynomial time and on which rank operators over \emph{every} prime 
field $\field F_p$, $p \neq q$ can be simulated in \FPC.
On the other hand, rank operators over $\field F_q$ can be used to canonise 
structures in $\mcK_q$ which means that the extension of fixed-point logic by
$\rkx{q}$-operators captures polynomial time on $\mcK_q$.
From this result we can easily extract the following answers to the open 
questions outlined above.
\begin{enumerate}[(a)]
 \item Rank logic (as defined in \cite{DaGrHoLa09}) fails to capture polynomial 
time (Theorem~\ref{thm:ranklogic:ptime}).
 \item The extensions of fixed-point logic by rank operators over 
different prime fields are incomparable (Theorem~\ref{thm:fpr:mainthm}), cf.\ 
\cite{Ho10,DaHo12,La11}.
\end{enumerate}

\smallskip
We obtain these classes of structures $\mcK_q$ by generalising the 
well-known construction of Cai, Fürer and Immerman~\cite{CFI92}.
It has been observed that their construction actually is a clever way of 
encoding a linear equation system over $\field F_2$ into an appropriate graph 
structure (see e.g.\ \cite{AtBuDa09, DaGrHoLa09, Ho10, La11}).
Intuitively, each gadget in the CFI-construction can be seen as an 
equation (or, equivalently, as a circuit gate) which counts the number of 
transpositions of adjacent edges modulo two, and the CFI-query is to decide 
whether the total number of such transpositions is even or odd.
Knowing this, it is very natural to ask whether this idea can be generalised to 
encode linear equation systems over arbitrary finite fields or, more generally, 
equation systems over arbitrary (Abelian) groups.

In~\cite{To04}, in order to obtain hardness results for 
the graph isomorphism problem, Tor\'{a}n followed this idea and established a 
graph construction which simulates $\modulo k$-counting gates
for all $k \geq 2$. 
Moreover, in order to separate the fragments of rank logic by operators over 
different prime fields, Holm presented in~\cite{Ho10} an even more general kind 
of construction which allows the representation of equations over \emph{every} 
Abelian group $G$. In fact, we obtain the classes $\mcK_q$ essentially by 
using his construction for the special case where $G = \field F_q$.

\subparagraph*{Solvability logic}
One important step in our proof is to consider \emph{solvability 
logic} \FPS which is the extension of \FPC by quantifiers which can express the 
solvability of linear equation systems over finite fields (so called 
\emph{solvability quantifiers}, see~\cite{DaGrHoKoPa13, Pa10}). 
Obviously the logic $\FPS$ can easily be embedded into rank logic (as rank 
operators can be used to solve linear equation systems), but it remains open 
whether the inclusion $\FPS \leq \FPR$ is strict.
To prove our main result outlined above we show that over certain classes of 
structures the logics $\FPS$ and $\FPR$ have precisely the same expressive 
power. In a more general context this might give some evidence that in the 
framework of fixed-point logic with counting rank operators can be simulated by 
solvability quantifiers.
On the other hand we show in Section~\ref{sec:slv:rk} that the 
extension of first-order logic (without counting) by solvability quantifiers 
is strictly weaker than the respective extension by rank operators.
This last result thus separates solvability quantifiers and rank 
operators in  the absence of counting.

\bigskip
Let us briefly sketch the main idea of our proofs which is to exploit the 
symmetries of definable linear equation systems. 
To this end, let $M \cdot \vct x = \onevec$ be a linear 
equation 
system over some prime field $\field F_p$ where $M$ is an $I \times 
I$-matrix over $\field F_p$ and where $\onevec$ is the $I$-identity vector over 
$\field F_p$, i.e.\ $\onevec(i) = 1$ for all $i \in I$.
Moreover, let $\Gamma$ be a group which acts on $I$ and which 
stabilises $M$, i.e.\ for all $i, j \in I$ and $\pi \in \Gamma$ we have 
$M(i,j) = M(\pi(i),\pi(j))$. 
In other words, if we identify the elements $\pi \in \Gamma$ with $I \times 
I$-permutation matrices $\Pi$ then we have $\Pi \cdot M = M \cdot \Pi$.
Now let $\vec b \in \field F_p^I$ be a solution of the linear 
equation system $M \cdot \vct x = \onevec$. Then we observe that also $\Pi 
\cdot \vct b$ is a solution for $\pi \in \Gamma$ since
\[ M \cdot (\Pi \cdot \vct b) = (M \cdot \Pi) \cdot \vct b = \Pi \cdot (M \cdot 
\vct b) = \Pi \cdot \onevec = \onevec.\]
In other words: the solution space of the linear equation system $M \cdot \vct 
x = \onevec$ is closed under the action of $\Gamma$.
Such and similar observations will enable us to transform a given 
linear equation system into a considerably simpler linear system which still is 
equivalent to the original one.

\section{Logics with linear-algebraic operators}
\label{sec:logics:linalgop}

By $\Str(\tau)$ we denote the class of all \emph{finite, relational} structures 
of signature $\tau$.
We assume that the reader is familiar with \emph{first-order logic} ($\FO$) and 
\emph{inflationary fixed-point logic} ($\FP$). For details 
see~\cite{ebbinghaus99finite,FMTbook}.
We write $\bbP$ for the set of primes and denote the prime field with $p$ 
elements by $\field F_p$. 
We consider matrices and 
vectors over \emph{unordered} index sets. Formally, if 
$I$ and $J$ are non-empty sets, then an $I \times J$-matrix $M$ over $\field 
F_p$ is a mapping $M: I \times J \to \field F_p$ and an $I$-vector $\vct v$ 
over $\field F_p$ is a mapping $\vct v: I \mapsto \field F_p$.

A \emph{(linear) preorder} $\preceq \,\, \subseteq A\times A$ on $A$ is a 
reflexive, transitive and total binary relation. A preorder $\preceq$ induces a 
linear order on the classes of the associated equivalence relation $x \sim y := 
(x \preceq y \wedge y \preceq x)$. 
We write $A = C_0 \preceq \cdots \preceq C_{n-1}$ to denote the decomposition 
of $A$ into $\sim$-classes $C_i$ which are linearly ordered by $\preceq$ as 
indicated.

We briefly recall the definitions of \emph{first-order logic with 
counting} $\FOC$ and \emph{(inflationary) fixed-point logic with counting} 
$\FPC$ which are the extensions of $\FO$ and $\FP$ by counting terms.
Formulas of $\FOC$ and $\FPC$ are evaluated over 
the \emph{two-sorted extension} of an input structure by a copy of the 
arithmetic. Following~\cite{DaGrHoLa09} we let $\mfA^{\#}$ denote the 
two-sorted extension of a $\tau$-structure $\mfA=(A,R_1,\dots,R_k)$ by the 
arithmetic $\mfN=(\N,+,\cdot,0,1)$, i.e.\ the two-sorted structure $\mfA^{\#} = 
( A, R_1,\dots , R_k , \N, +, \cdot, 0, 1)$ where the universe of the first 
sort (also referred to as \emph{vertex sort}) is $A$ and the universe of the 
second sort (also referred to as \emph{number sort} or \emph{counting sort}) is 
$\N$.

As usual for the two-sorted setting we have,
for both, the vertex and the number sort, a collection of typed
first-order variables. We agree to use Latin letters $x, y, z, \dots$
for variables which range over the vertices and Greek letters $\nu, \mu,
\dots$ for variables ranging over the numbers. 
Similarly, for second-order variables $R$ we allow mixed types, i.e.\ a 
relation symbol $R$ of type $(k , \ell) \in \N \times \N$ stands for a relation 
$R \subseteq A^k \times \N^\ell$. Of course, already first-order logic over 
such 
two-sorted extensions is undecidable. To obtain logics whose 
data complexity is in polynomial time we restrict the quantification over the 
number sort by a numeric term $t$, i.e.\ $Q\nu \leq t . \phi$ where $Q \in \{ 
\exists, \forall \}$ and where $t$ is a closed \emph{numeric} term.
Similarly, for fixed-point logic \FP we bound the numeric components of 
fixed-point variables $R$ of type $(k,\ell)$ in all fixed-point definitions 
\[
 \left[ \ifp \, R\bx\bnu\leq \bar{t} \,.\, (\phi(\bx,\bnu)) \right](\bx,\bnu)
\]
by a tuple of closed numeric terms $\bar{t}=(t_1,\dots,t_\ell)$ where each 
$t_i$ bounds the range of the variable $\nu_i$ in the tuple $\bnu$. 
For the logics which we consider here the value of such 
numeric terms (and thus the range of all quantifiers over the number sort) 
is polynomially bounded in the size of the input structure. Together 
with the standard argument that inflationary fixed-points can be evaluated in 
polynomial time and the fact that the matrix rank over any field can be 
determined in polynomial time (for example by the method of Gaussian 
elimination), this ensures that all the logics which we introduce in the 
following have polynomial-time data complexity.

Let $\bx\bnu$ be a mixed tuple of variables and let $\bar t$ be a tuple of 
closed numeric terms which bounds the range of the numeric variables in $\bnu$.
For a formula $\phi$ we define a \emph{counting term} $s = [\# \bx\bnu \leq 
\bar t \,.\, \phi]$ whose value $s^\mfA \in \N$ in a structure $\mfA$ 
corresponds to the number of tuples 
$(\ba,\bar n) \in A^k \times \N^\ell$ such that $\mfA \models \phi(\ba, \bar 
n)$ 
and $n_i \leq t_i^\mfA$  where $k = \card \bx$ and $\ell = \card {\bnu }$.

We define \emph{first-order logic with counting $\FOC$} as the extension of 
(the above described two-sorted variant of) $\FO$ by counting terms. 
Similarly, by adding counting terms to the logic $\FP$ we obtain
\emph{(inflationary) fixed-point logic with counting $\FPC$}.

\paragraph*{Extensions by rank operators}

Next, we recall the notion of rank operators as introduced 
in~\cite{DaGrHoLa09}. 
Let $\Theta(\bx\bnu\leq \bar t ,\by\bmu \leq \bar s)$ be a numeric 
term where $\bar t$ and $\bar s$ are tuples of closed numeric terms which bound 
the 
range of the numeric variables in the tuples $\bnu$ and $\bmu$, respectively. 
Given a structure $\mfA$ we define $\N^{\leq \bar t} \defeq \{ \bn \in 
\N^{\card \bnu} : n_i \leq t_i^\mfA \}$. 
The set $\N^{\leq \bar s} \subset \N^{\card \bmu}$ is defined 
analogously. The term $\Theta$ defines in the structure $\mfA$ 
for $I \defeq A^{\card \bx} \times \N^{\leq \bar t}$ and $J \defeq A^{\card 
\by} \times \N^{\leq \bar s}$
the $I \times J$-matrix $M_\Theta$ with values in $\N$ that is given
as $M_\Theta(\ba\bn, \bb\bm) \defeq \Theta^\mfA(\ba\bn,\bb\bm)$.

The \emph{matrix rank operators} compute the rank of the matrix $M_\Theta$ 
over a prime field $\field F_p$ for $p \in \bbP$.
Firstly, as in~\cite{DaGrHoLa09}, we define for every prime $p$ a matrix rank 
operator $\rkp$ which allows us to construct a new numeric \emph{rank term} $[ 
\rkp 
\,(\bx\bnu\leq 
\bar t ,\by\bmu \leq \bar s) \, . \, \Theta ]$ whose value in the 
structure $\mfA$ is the rank of the matrix $(M_\Theta \,\,\modulo p)$ over 
$\field 
F_p$.
Secondly, we propose a more flexible rank operator $\rk$ which gets the prime 
$p$ as an additional input. Formally, with this rank operator $\rk$ we can 
construct a rank term $[ \rk \,(\bx\bnu\leq \bar t ,\by\bmu \leq \bar s,\pi 
\leq r) \, . \, \Theta ]$ where $\pi$ is an additional free numeric variable 
whose range is bounded by some closed numeric term $r$. Given a structure 
$\mfA$ 
and an assignment $\pi \mapsto p$ for some prime $p \leq r^\mfA$, the value of 
this rank term is the matrix rank of $(M_\Theta \,\,\modulo p)$ considered as a 
matrix over $\field F_p$.
The rank operator $\rk$ can be seen as a unification for the 
the family of rank operators $(\rkp)_{p \in \bbP}$ and has been introduced 
in~\cite{Ho10, La11, Pa10}.

\smallskip
We define, for every set of primes $\Omega \subseteq \bbP$, the extension 
$\FORx{\Omega}$ of $\FOC$ and the extension $\FPRx{\Omega}$ of $\FPC$ by  
matrix rank operators $\rkp$ with $p \in \Omega$. 
For convenience, we let $\FOR = \FORx{\bbP}$ and $\FPR = \FPRx{\bbP}$.
Similarly, we denote by $\FPRvar$ the extension of $\FPC$ by the uniform rank 
operator $\rk$. 
We remark, that rank operators can directly simulate counting terms. For 
example we have that 
\[ [\# x \,.\, \phi(x) ] = [ \rkp \,(x ,y) \, . \, ( x = y \wedge \phi(x)) ]. 
\] 
Hence, we could equivalently define the rank logics $\FORx{\Omega}, 
\FPRx{\Omega}$ and $\FPRvar$ as the extensions of (the two-sorted variants of) 
$\FO$ and $\FP$, respectively.
\smallskip

\paragraph*{Extensions by solvability quantifiers}

It is well-known that the extensions of $\FOC$ and $\FPC$ by matrix rank 
operators have surprising expressive power which, in particular, goes beyond 
that of fixed-point logic with counting. 
For example, it is known that rank operators can easily define the 
symmetric transitive closure of binary relations and that they can be used 
to express the structure isomorphism problem on various classes on which
the Weisfeiler-Lehman test fails like, for example, classes of Cai, Fürer and 
Immerman graphs~\cite{CFI92, DaGrHoLa09}.
Interestingly, such results for rank logic were obtained by reducing the 
respective queries to a \emph{solvability problem for linear equation system 
over finite fields}.
Although the solvability problem (for linear equation systems) can 
be defined in rank logic, we propose to study extensions by 
quantifiers which directly express this solvability problem. One 
advantage of this approach is that one can naturally define such 
quantifiers for 
linear systems over more 
general classes of algebraic domains, like rings, for which no appropriate 
notion of matrix rank exists, cf.\ \cite{DaGrHoKoPa13}. 

\medskip
Let $\Omega \subseteq \bbP$ be a set of primes. Then the \emph{solvability 
logic} $\FPSx{\Omega}$ extends the syntax of $\FPC$ for every $p \in 
\Omega$ by the following formula creation rule for \emph{solvability 
quantifiers} $\slvp$.
\begin{itemize}
 \item Let $\phi(\bx\bnu,\by\bmu,\bz) \in \FPSx{\Omega}$ and let 
 $\bar t$ and $\bar s$ be tuples of closed numeric terms with $\card {\bar 
t 
\,} = \card{\bnu}$ and $\card { \bar s \, } = \card{\bmu}$.
  Then also $\psi(\bz) = (\slvp \, \bx\bnu\leq \bar s, \by\bmu \leq \bar 
t) 
\phi(\bx\bnu,\by\bmu,\bz)$ is a formula of $\FPSx{\Omega}$. 
\end{itemize}

The semantics of the formula $\psi(\bz)$ is defined similarly as for rank logic.
More precisely, let $k = \card{\bx}$ and $\ell = \card{\by}$. 
To a pair $(\mfA, \bz \mapsto \bc) \in \Str(\sigma, \bz)$ we 
associate the $I\times J$-matrix $M_\phi$ over $\{ 0, 1 \}
\subseteq \field F_p$ where $I = A^k \times \N^{\leq \bar s}$ and $J = 
A^\ell \times \N^{\leq \bar t}$ and where for $\ba \in I$ and $\bb \in J$ 
we have $M_\phi(\ba,\bb) = 1$ if, and only if, $\mfA \models 
\phi(\ba,\bb,\bc)$.

Let $\onevec$ be the $I$-identity vector over $\field F_p$, i.e.\ 
$\onevec(\ba) = 1$ for all $\ba \in I$. Then $M_\phi$ and $\onevec$ 
determine
the linear
equation system $M_\phi \cdot \vct x = \onevec$ over $\field F_p$ where 
$\vct x
= (x_j)_{j \in J}$ is a $J$-vector of variables $x_j$ which range over 
$\field
F_p$. Finally, $\mfA \models \psi(\bc)$ if, and only if, $M_\phi 
\cdot \vct x = \onevec$ is solvable.

\smallskip
At first glance, the solvability quantifier seem to pose 
serious restrictions on 
the syntactic form of definable linear equation systems.
Specifically, the coefficient matrix has to be a matrix over $\{0,1\}$ and
the vector of constants is fixed from outside.
However, it is not hard to show that general linear equation systems can 
be brought into this kind of normal form by using quantifier-free 
first-order transformations (see Lemma~4.1 in~\cite{DaGrHoKoPa13}).

We write $\FPS$ to denote the logic $\FPSx{\Primes}$ and $\FPSp$ to denote 
the logic $\FPSx{\{ p \}}$ for $p \in \bbP$. 
Analogously to the definition of $\FPRvar$ we also consider a 
solvability quantifier $\slv$ which gets the prime $p$ as an 
additional input and which can uniformly simulate all solvability 
quantifiers $\slvp$ for $p \in \bbP$. 
Let $\FPSvar$ denote the extension of $\FPC$ by this uniform version of a 
solvability quantifier.
Then the following inclusions easily follow from the definitions and the fact 
that rank operators can be used to define the solvability problem for linear 
equation systems.
\[
   \begin{array}{ccccccccc}
    \FOR_p & \leq & \FPRp & \leq & \FPR  & \leq & \FPRvar & \leq & \PTIME \\
    \text{\rotatebox{90}{$\leq$}}  &  &   \text{\rotatebox{90}{$\leq$}} &  &
    \text{\rotatebox{90}{$\leq$}} & &   \text{\rotatebox{90}{$\leq$}} & & \\
    \FOSp & \leq & \FPSp & \leq & \FPS  & \leq & \FPSvar & & \\
   &  &   \text{\rotatebox{90}{$\leq$}} &  & & &    & & \\
     &  & \FPC & &   &  &  & & \\
  \end{array}
\]

Finally we remark that, analogously to \cite{DaGrHoLa09}, we defined rank 
operators and solvability quantifiers for prime fields only. Of course, the 
definition can easily be generalised to cover all finite fields, i.e.\ also 
finite fields of prime power order. 
However, for the case of solvability quantifiers, Holm was able to prove 
in~\cite{Ho10} that this does not alter the expressive power of the resulting 
logics since solvability quantifiers over a finite field $\field F_q$ of prime 
power order $q = p^k$ can be simulated by solvability quantifiers over $\field 
F_p$. Moreover, a similar reduction can be achieved for rank operators which 
altogether shows that it suffices to focus on rank operators and solvability 
quantifiers over prime fields.

\section{Separation results over different classes of fields}
\label{sec:sep:fields}

In this section we separate the extensions of fixed-point logic with counting 
by solvability quantifiers and rank operators over different prime fields. 
Specifically, we show that the expressive power of the logics $\FPSx{\Omega}$ 
is different for all sets of primes $\Omega$.
Moreover, we transfer these results to the extensions $\FPRx{\Omega}$ by rank 
operators. In this way we can answer the following open question about rank 
logic:
\begin{center}
 It holds that $\FPRx{p} \neq \FPRx{q}$ for pairs of different primes $p,q$. 
\cite{DaHo12, Ho10, La11}
\end{center}

Another important consequence of our result is that rank logic (in the way it 
was defined in~\cite{DaGrHoLa09}) does not suffice to capture 
polynomial time. 
Let us state these results formally.

\begin{theorem}\label{thm:fpr:mainthm}
 Let $\Omega \neq \Omega'$ be two sets of primes.
Then $\FPSx{\Omega} \neq \FPSx{\Omega'}$ and $\FPRx{\Omega} \neq 
\FPRx{\Omega'}$.
\end{theorem}

\begin{theorem} \label{thm:ranklogic:ptime}
Rank logic fails to capture polynomial time. We have $\FPR < \FPRvar \leq 
\PTIME$.
\end{theorem}

\noindent
In fact, both theorems are simple consequences of our following main result.
\begin{theorem}\label{thm:fps:distinct:primes}
For every prime $q$ there is a class of structures $\mcK_q$ such that
\begin{enumerate}[(a)]
  \item $\FPSx{\Omega} = \FPC$ on $\mcK_q$ for every set of primes $\Omega$ 
with $q \nin \Omega$, 
\label{thm:fps:distinct:primes:fps:fpc}
 \item $\FPRx{\Omega} = \FPSx{\Omega}$ on $\mcK_q$ for all sets of primes 
$\Omega$, 
 \label{thm:fps:distinct:primes:fpr:fps}
 \item $\FPC < \PTIME$ on $\mcK_q$, and
 \label{thm:fps:distinct:primes:fpc:not:ptime}
  \item $\FPSx{q}=\PTIME$ on $\mcK_q$.
 \label{thm:fps:fpsq:ptime}
\end{enumerate} 
\end{theorem}

\begin{proof}[Proof of Theorem~\ref{thm:fpr:mainthm}]
 Let $\Omega$ and $\Omega'$ be two sets of primes as above. Without loss of 
generality let us assume that there exists a prime $q \in \Omega \setminus 
\Omega'$. Then by Theorem~\ref{thm:fps:distinct:primes} there exists a class 
$\mcK_q$ on which $\FPSx{\Omega} = \FPRx{\Omega} = \PTIME$ and on which
$\FPSx{\Omega'} = \FPRx{\Omega'} = \FPC < \PTIME$.
\end{proof}

\begin{proof}[Proof of Theorem~\ref{thm:ranklogic:ptime}]
 Otherwise assume that $\FPR = \PTIME$. 
Then, in particular, \FPR = \FPRvar and there exists a formula $\phi \in 
\FPR$ which can uniformly determine the rank of matrices over prime fields, 
i.e.\ which can express the uniform rank operator $\rk$.
As a matter of fact we have $\phi \in \FPRx{\Omega}$ for some \emph{finite} set 
of primes $\Omega$.
By using $\phi$ we can uniformly express the matrix rank over each prime field 
$\field F_p$ in $\FPRx{\Omega}$. In other words, we have $\FPS \leq \FPR \leq 
\FPRvar \leq \FPRx{\Omega}$.

Now let $q \in \Primes \setminus \Omega$. 
By Theorem~\ref{thm:fps:distinct:primes} there exists a class of 
structures $\mcK_q$ on which $\FPRx{\Omega} = \FPC < \PTIME$.
However, the class $\mcK_q$ can be chosen such that $\PTIME = 
\FPSx q \leq \FPRx{\Omega}$ on $\mcK_q$ by
Theorem~\ref{thm:fps:distinct:primes}~(\ref{thm:fps:fpsq:ptime})
and we obtain the desired contradiction.
\end{proof}

The proof of Theorem~\ref{thm:ranklogic:ptime} reveals a deficiency of the 
logic \FPR: each formula can only access $\rkp$-operators for a finite 
set $\Omega$ of distinct primes $p$. In fact, the query which we constructed to 
separate $\FPR$ from $\PTIME$ can be defined in $\FPRvar$. 
Altogether this suggests to generalise the notion of rank operators 
and to specify the prime $p$ as a part of the input, as we did for $\FPRvar$, 
and as it was proposed in \cite{Ho10, La11, Pa10}.

\medskip
The remainder of this section is devoted to the proof 
of Theorem~\ref{thm:fps:distinct:primes}.
We fix a prime $q$ and proceed as follows.
In a first step, we identify properties of classes of 
structures $\mcK_q$ which guarantee that the relations claimed in
(\ref{thm:fps:distinct:primes:fps:fpc}),
(\ref{thm:fps:distinct:primes:fpr:fps}), 
(\ref{thm:fps:distinct:primes:fpc:not:ptime}) and
(\ref{thm:fps:fpsq:ptime}) hold.
In a second step, we proceed to show that we can obtain a
class of structures $\mcK_q$ that satisfies all of these sufficient 
criteria. This together then proves our theorem.

\paragraph*{Establishing sufficient criteria}
We start by establishing sufficient criteria for the most relevant 
part of Theorem~\ref{thm:fps:distinct:primes}
which is the relation claimed in~(\ref{thm:fps:distinct:primes:fps:fpc}).
Assume that we have a class of structures $\mcK_q = 
\mcK$ with the following properties.
\begin{enumerate}[$(I)$]
 \item \label{item:property:classk:autogroups}
 The automorphism groups $\Delta_\mfA \defeq \Aut(\mfA)$ of 
structures $\mfA \in \mcK$ are Abelian $q$-groups.
 \item \label{item:property:classk:deforbits}
 The orbits of $\ell$-tuples in structures $\mfA \in \mcK$ can be ordered in 
\FPC.

Formally, for each $\ell \geq 1$ there is a formula 
$\phi_\preceq(x_1,\dots,x_\ell, y_1, \dots, y_\ell) \in \FPC$ such that for 
every structure $\mfA \in \mcK$, the formula $\phi_\preceq(\bx,\by)$ defines in 
 $\mfA$ a linear preorder $\preceq$ on $A^\ell$ with the property that two 
$\ell$-tuples $\ba, \bb \in A^\ell$ 
are $\preceq$-equivalent if, and only if, they are in the same 
$\Delta_\mfA$-orbit.
\end{enumerate}

\begin{lemma}
\label{lemma:fps:fpc:prop12}
 If $\mcK$ satisfies (\ref{item:property:classk:autogroups}) and
 (\ref{item:property:classk:deforbits}), then $\FPSx{\Omega} = \FPC$ on $\mcK$ 
for all $\Omega \subseteq \bbP\setminus \{q\}$.
\end{lemma}

The proof of this lemma is by induction on the structure of 
$\FPSx{\Omega}$-formulas. 
Obviously, 
the only interesting step is the translation of a solvability formula
\[  \psi(\bz) = (\slvp \, \bx\bnu\leq \bar s, \by\bmu \leq \bar t) 
\phi(\bx\bnu,\by\bmu,\bz)
\]
into an $\FPC$-formula $\vartheta(\bz)$ which is equivalent to 
$\psi(\bz)$ on the class $\mcK$.
Let $\card \bx = \card \by = \ell$, $\card \bnu = \card \bmu = \lambda$ 
and 
$\card \bz = k$. To explain our main argument, we fix a structure $\mfA 
\in \mcK$ and a $k$-tuple of parameters $\bc \in (A \uplus \N )^k$ which 
is compatible with the type of the variable tuple $\bz$. 
According to the semantics of the solvability quantifier, the formula 
$\phi$ defines in 
$(\mfA, \bz \mapsto \bc)$ an $I \times J$-matrix $M = M^{\mfA}_{\bc}$ with 
entries in $\{0,1\} \subseteq \field F_p$ where  $I = I^{\mfA}_{\bc} 
\defeq A^\ell \times {\N}^{\leq \bar s} \subseteq A^\ell \times 
{\N}^\lambda $
and $J = J^{\mfA}_{\bc} \defeq A^\ell \times {\N}^{\leq \bar t} \subseteq 
A^\ell \times 
{\N}^\lambda$ that is defined for $\ba \in  I$ and $\bb \in J$ as
\[ M(\ba,\bb) = \begin{cases}
                 1, &\text{ if } \mfA \models \phi(\ba,\bb,\bc)\\
                 0, &\text{ else.}
                \end{cases}\]

By definition we have $\mfA \models \psi(\bc)$ if, and only if, the linear 
equation system $M 
\cdot \vec x = \onevec$ over $\field F_p$ is solvable.
The key idea is to use the symmetries of the structure $\mfA$ to translate 
the linear equation system $M \cdot \vec x = \onevec$ into 
an equivalent linear system which is \emph{simpler} in the sense that its
solvability can be defined in the logic $\FPC$.
The reader should observe that each automorphism $\pi \in 
\Delta_\mfA = \Aut(\mfA)$ naturally 
induces an automorphism of the two-sorted extension $\mfA^{\#}$ which 
point-wise fixes every number $n \in \N$. In particular we 
have $\Aut(\mfA) = \Aut(\mfA^{\#})$.

\medskip
We set $\Gamma = \Gamma_\bc^\mfA \defeq 
\Aut(\mfA,\bc) \leq \Delta = \Delta_\mfA = \Aut(\mfA)$. 
The group $\Gamma$ acts on $I$ and $J$ in the natural way. We identify 
each automorphism $\pi \in \Gamma$ with the corresponding $I\times 
I$-permutation matrix $\Pi_I$ and the corresponding $J \times J$-permutation 
matrix $\Pi_J$ in the usual way. More precisely, to $\pi \in \Gamma$ 
we associate the $I \times I$-permutation matrix $\Pi_I$ which is defined as
\[ \Pi_I(\bar a, \bar b) = \begin{cases}
                            1,&\pi(\bar a) = \bar b \\
                            0,&\text{otherwise}.
                           \end{cases}\]
Then $\Gamma$ acts on the set of $I \times J$-matrices by left
multiplication with $I \times I$-permutation matrices. Similarly, we let
$\Pi_J$ denote the $J \times J$-permutation matrix defined as 
\[ \Pi_J(\bar a, \bar b) = \begin{cases}
                            1,&\pi(\bar a) = \bar b \\
                            0,&\text{otherwise}.
                           \end{cases}\]
Then $\Gamma$ also acts on the set of $I \times J$-matrices by right
multiplication with $J \times J$-permutation matrices. Specifically, for
$\pi \in \Gamma$ we have $(\Pi_I \cdot M)(\bar a, \bar b) = M(\pi(\bar
a), \bar b)$ and $(M \cdot \Pi_J^{-1})(\bar a, \bar b) = M(\bar a, \pi(\bar
b))$. Since $M$ is defined by a formula in the structure $(\mfA, \bc)$ and 
since $\Gamma = \Aut(\mfA, \bc)$ we conclude that $(\Pi_I \cdot M \cdot 
\Pi_J^{-1})(\bar a, \bar b) = M(\pi(\bar
a), \pi(\bar b)) = M(\bar a, \bar b)$ and thus
\[ \Pi_I \cdot M \cdot \Pi_J^{-1} = M \quad\Leftrightarrow\quad \Pi_I \cdot
M = M \cdot \Pi_J. \]
This identity leads to the following important observation.

\begin{lemma}\label{lemma:fps:symmetric:solution}
 If $M \cdot \vec x = \onevec$ is solvable, then 
the system has a \emph{$\Gamma$-symmetric} solution, i.e.\ a solution $\vec 
b \in 
\field F_p^J$ such that $\Pi_J \cdot \vec b = \vec b$ for all $\pi \in 
\Gamma$.
\end{lemma}
\begin{proof}
If $M \cdot \vec b = \onevec$, then also $\Pi_I \cdot (M \cdot 
\vec b ) = \onevec$ and thus $M \cdot (\Pi_J \cdot \vec b ) = \onevec$ for all 
$\pi \in \Gamma$. This shows that $\Gamma$ acts on the solution space of the 
linear equation system. 
Since $\mcK$ satisfies property~(\ref{item:property:classk:autogroups}) we know 
that $\Gamma$ is a $q$-group for a prime $q \neq p$. Thus each $\Gamma$-orbit 
has size $q^r$ for some $r \geq 0$. On the other hand, the number of 
solutions is a power of $p$. We conclude that there is at least one 
$\Gamma$-orbit of size one which proves our claim.
\end{proof}

Let $\vec b \in \field F_p^J$ be a $\Gamma$-symmetric solution. Then the 
entries of the solution $\vec b$ on $\Gamma$-orbits are constant: for $j \in J$ 
and $\pi \in \Gamma$ we have $\vec b(\pi(j)) = (\Pi_J \cdot \vec b)(j) = \vec 
b(j)$.
We proceed to use the property (\ref{item:property:classk:deforbits}) and 
show that there exists an \FPC-formula $\phi_\preceq(\bx,\by)$  which 
defines for all $\mfA \in \mcK$ and $\bc \in (A\uplus \N)^k$ as above a 
linear preorder $\preceq$ on $A^\ell$ which identifies $\Gamma$-orbits. 
Note that, in general, $\Gamma=\Aut(\mfA,\bc)$ is a strict subgroup of 
$\Delta=\Aut(\mfA)$. Thus we can not directly apply 
(\ref{item:property:classk:deforbits}).
However, the $\Gamma$-orbits on 
$A^\ell$ correspond to the $\Delta$-orbits on 
$A^{k'+\ell}$ where the first $k'$ entries are fixed to the elements $\{ 
c_1, \dots, c_k \} \cap A$.

The linear preorder $\preceq$ naturally extends to a preorder on the 
sets $I$ and $J$ with the same properties.
Let us write $J = J_0 \preceq J_1 \preceq \cdots \preceq J_{v-1}$ to denote the 
decomposition of $J$ into the $\Gamma$-orbits $J_j$ which are ordered 
by $\preceq$ as indicated.
Moreover, for $j \in \inseg v$ we let $e_j$ denote the identity vector on the 
$j$-th orbit $J_j$, i.e.\ the $J$-vector which defined for $i \in J$ as
\[ e_j(i) \defeq 
\begin{cases}
1,&\text{ if } i \in J_j \\
0,&\text{ else.}
\end{cases} \]

Let $E$ denote the $J \times \inseg v$-matrix whose $j$-th column is the vector 
$e_j$. It follows that a $\Gamma$-symmetric solution $\vec b$ can be written as 
$E \cdot \vec b_* = \vec b$ for a unique $\inseg v$-vector $\vec b_*$.
Together with Lemma~\ref{lemma:fps:symmetric:solution} this shows the following.

\begin{lemma}
 The linear equation system $M \cdot \vec x = \onevec$ is solvable if, and only 
if, the linear equation system $(M \cdot E) \cdot \vec x_* = \onevec$ is 
solvable.
\end{lemma}

Finally, we observe that the coefficient matrix $M_* \defeq (M 
\cdot E)$ of the equivalent linear equation system $M_* \cdot \vec x_* = 
\onevec$ can easily be obtained in $\FPC$ and that it is a matrix over the 
\emph{ordered} set of column indices $\inseg v$. 
It is a simple observation that such linear equation systems can be solved 
in $\FPC$: the linear order on the column set induces 
(together with some fixed order on $\field F_p$) a lexicographical 
ordering on the set of rows which is, up to duplicates of rows, a 
linear order on this set. Thus, in general, if we have a linear order on 
\emph{one} of the index sets of the coefficient matrix this suffices to 
obtain an equivalent matrix where \emph{both} index sets are ordered, see 
also~\cite{Pa10}.
This finishes our proof of Lemma~\ref{lemma:fps:fpc:prop12}.

\bigskip
We proceed to show that the conditions 
(\ref{item:property:classk:autogroups}) and
 (\ref{item:property:classk:deforbits}) also  guarantee that 
rank operators can be reduced to solvability operators over the class 
$\mcK$. In fact, for this translation we only require the somewhat weaker 
assumption that we can define in \FPC on $\ell$-tuples in 
structures $\mfA \in \mcK$ a linear preorder in which every class can be 
linearised in \FPC by fixing a constant number of parameters. 
The precise technical requirements will become clear from the proof of the 
following lemma.
\begin{lemma}\label{lemma:rank:to:solve}
If $\mcK$ satisfies (\ref{item:property:classk:autogroups})
 and (\ref{item:property:classk:deforbits}), then
 $\FPRx{\Omega} = \FPSx{\Omega}$ on $\mcK$ for all sets of primes $\Omega$.
\end{lemma}
\begin{proof}
 We inductively translate $\FPRx{\Omega}$-formulas into formulas of 
$\FPSx{\Omega}$ which 
are equivalent on the class $\mcK$. The only interesting case is 
the transformation of rank terms
\[ \Upsilon(\bz) = [ \rkp \,(\bx\bnu\leq \bar t ,\by\bmu \leq \bar s) \, . 
\, \Theta(\bx\bnu, \by\bmu, \bz) ]. \]

Let $\card \bx = \card \by = \ell$, $\card \bnu = \card \bmu = \lambda$ 
and 
$\card \bz = k$. Let $\mfA \in \mcK$ and let $\bc$ be a $k$-tuple of 
parameters $\bc \in (A \uplus \N )^k$ which 
is compatible with the type of the variable tuple $\bz$. 
The term $\Theta$ defines in $(\mfA, \bz \mapsto \bc)$ 
for $I^{\mfA} = I \defeq A^{\card \bx} \times \N^{\leq \bar t}$ and 
$J^{\mfA} = J \defeq A^{\card \by} \times \N^{\leq \bar s}$
the $I \times J$-matrix $M$ over $\field F_p$ which is defined as
\[ M(\ba\bn, \bb\bm) \defeq \Theta^\mfA(\ba\bn,\bb\bm,\bc) \,\,\modulo p.\]

According to the semantics of matrix rank operators, the 
value $\Upsilon^\mfA(\bc) \in \N$ is the rank of the matrix $M$.
We proceed to show that we can determine the matrix rank of $M$ by a 
recursive application of solvability queries. To this end we make the 
following key observation.

\medskip
\begin{claim}
 There are $\FPC$-formulas $\phi_\preceq(\by_1\bmu_1,\by_2\bmu_2)$, 
$\psi_\leq(\bv,\by_1\bmu_1,\by_2\bmu_2)$ such that for every 
$\mfA \in \mcK$
\begin{enumerate}[(a)]
 \item $\phi_\preceq^\mfA$ is a linear preorder $\preceq$ on 
$J^\mfA$, and such that
 \item for every $\preceq$-class $[j] \subseteq J^\mfA$ there exists a 
parameter tuple $\bd \in A^{\card \bv}$ such that $\psi_\leq^\mfA(\bd)$ is 
a linear order $\leq$ on $[j]$.
\end{enumerate}
\end{claim}
\begin{claimproof}
 First of all, we let $\phi_\preceq$ be an \FPC-formula which defines in 
every structure $\mfA \in \mcK$ a linear preorder $\preceq$ on $J^\mfA$ 
such that $\preceq$-classes correspond to $\Delta_\mfA$-orbits. Such a 
formula exists by our assumption that $\mcK$ satisfies 
property~(\ref{item:property:classk:deforbits}).
Analogously, we choose an $\FPC$-formula $\vartheta_\preceq$ which defines 
in every structure $\mfA \in \mcK$ a linear preorder $\preceq^*$ on 
$J^\mfA \times J^\mfA$ that induces a linear order on the 
$\Delta_\mfA$-orbits.

Now let $[j] \subseteq J^\mfA$ be a $\preceq$-class for some $\mfA \in 
\mcK$. 
By property (\ref{item:property:classk:autogroups}) we know that 
$\Delta_\mfA$ is an Abelian group. Thus, each automorphism $\pi \in 
\Delta_\mfA$ which fixes \emph{one} element in the 
$\Delta_\mfA$-orbit $[j]$ point-wise 
fixes \emph{every} element in the class $[j]$. 
We conclude that the restriction of $\preceq^*$ to elements in $\{j \} 
\times [j]$ corresponds to a linear order on $[j]$ for each $j \in [j]$.
In this way we obtain an \FPC-formula $\psi_\leq$ with the desired 
properties.
\end{claimproof}

\smallskip
We are now prepared to describe the recursive procedure which allows us to 
determine the rank of the matrix $M$ in $\FPSx{\Omega}$.
To this end we fix formulas $\phi_\preceq$ and $\psi_\leq$ with the above 
properties. Moreover, let $\preceq$ denote the linear preorder defined by 
$\phi_\preceq$ on $J$ and let
\[ J = J_0 \preceq J_1 \preceq \cdots \preceq J_{r-1}.\]
We use the formula $\psi_\leq$ to obtain on each class $J_i$ a 
family of definable linear orderings (which depend on the choice of 
different parameters).
For $j \in J$ we denote by $\vct{m}_j \in \field F_p^I$ the $j$-th column of 
the matrix $M$. Then the rank of $M$ coincides with the dimension of the 
$\field F_p$-vector space which is generated by the set of columns $\{ \vct m_j 
: j \in J\}$ of the matrix $M$.

Now, for $i \in \inseg r$ we recursively obtain the 
dimension $d_i \in \N$ of the $\field F_p$-vector space generated by 
$V_i \defeq \{ \vct m_j : j \in J_0 \cup J_1 \cup \cdots \cup J_i\}$ as 
follows.
First, we use $\psi_\leq$ to fix a linear order on $J_i$ (the following 
steps are independent of the specific linear order and can thus be 
performed in parallel for each such order).
Using this linear order on $J_i$ we can identify in $\FPSx{\Omega}$ a maximal 
set 
$W \subseteq \{ \vec m_j : j \in J_i\}$ of linearly independent columns 
such that $\gengroup {V_{i-1}} \cap \gengroup W = \{ \zerovec \}$.
Indeed, if $\gengroup {V_{i-1}} \cap \gengroup W = \{ \zerovec \}$, then 
for $\vec m \in \{ \vec m_j : j \in J_i\}$, $\vec m \nin \gengroup W$
we have that 
$\gengroup {V_{i-1}} \cap \gengroup {W \uplus \{ \vec m \}} = \{ \zerovec 
\}$ if, and only if, $\vec m \nin \gengroup { V_{i-1} \cup W}$. Observe 
that the conditions $\vec m \nin \gengroup W$ and $\vec m \nin \gengroup { 
V_{i-1} \cup W}$ correspond to the solvability of a linear equation system 
over $\field F_p$.
We claim that $d_i = d_{i-1} + \card W$. Indeed, by the 
maximality of $W$ and since $\gengroup{ V_{i-1}} \cap \gengroup W = \{ 
\zerovec \}$ it follows that $\gengroup {V_i} = \gengroup {V_{i-1} } 
\oplus \gengroup W$. Moreover, $W$ consists of linearly independent 
columns and is a basis for $\gengroup W$.

Since the above described recursion can easily be implemented in 
$\FPSx{\Omega}$, we conclude that the rank $d_{r-1}$ of the matrix $M$ can be 
determined in $\FPSx{\Omega}$ which completes our proof.
\end{proof}

\bigskip
We now focus on the parts 
(\ref{thm:fps:distinct:primes:fpc:not:ptime})
and 
(\ref{thm:fps:fpsq:ptime})
of Theorem~\ref{thm:fps:distinct:primes} and establish sufficient criteria 
which guarantee that $\FPC$ fails to capture $\PTIME$ on 
$\mcK$ while $\FPSx q$ can express every polynomial-time decidable 
property of $\mcK$-structures.

\begin{enumerate}[$(I)$]
\setcounter{enumi}{2}
 \item There exists an $\FPSx q$-definable canonisation procedure on $\mcK$.
\label{item:property:classk:isopoly}
 \item For every $k \geq 1$ there exists a pair of structures $\mfA \in 
\mcK$ and $\mfB \in \mcK$ such that $\mfA \nisom \mfB$ and $\mfA \Ckeqv 
\mfB$.
\label{item:property:classk:ckeqv}
\end{enumerate}
\begin{lemma}\label{lemma:fps:fpc:prop34}
 If $\mcK$ satisfies  (\ref{item:property:classk:isopoly}) and
 (\ref{item:property:classk:ckeqv}), then $\FPC < \FPSx q = \PTIME$ on~$\mcK$.
\end{lemma}
\begin{proof}
It is clear that by property (\ref{item:property:classk:isopoly}) we have
 $\FPSx q = \PTIME$ on $\mcK$.
 Moreover, if we had $\FPC = \PTIME$ on $\mcK$ then,  by the embedding of 
$\FPC$ 
into $\INFCkx{\omega}$ and the fact that $\mcK$-structures can be canonised in 
polynomial time, there exists a fixed $k\geq 1$ such that $\INFCk$ can identify 
each structure in $\mcK$ which, in turn, contradicts 
property~(\ref{item:property:classk:ckeqv}).
\end{proof}

\paragraph*{Constructing an appropriate class of structures}
We proceed to construct a class of structures $\mcK$ which satisfies 
properties (\ref{item:property:classk:autogroups}) - 
(\ref{item:property:classk:ckeqv}).
Our approach is a generalisation of the well-known construction of Cai, Fürer 
and Immerman~\cite{CFI92} for cyclic groups other than $\field F_2$. 
To illustrate the main differences, let us briefly recall the idea of the 
original construction. Starting from an undirected and connected graph 
$\mcG=(V,E)$, we take two copies $e_0, e_1$ of every edge $e \in E$.
Moreover, for every vertex $v \in V$ we consider the set $vE \subseteq E$ of 
edges which are adjacent to $v$ and we add one of the following two constraints 
to restrict the symmetries of the resulting CFI-graph: either the set of all 
sets $\{ 
e_{\rho(e)} : e \in vE\}$ with $\rho: vE 
\to \field F_2$ and $\sum_{e \in vE} \rho(e) = 0$ is stabilised (an \emph{even} 
node) or the dual set of all sets $\{ e_{\rho(e)} : e  \in vE\}$ with 
$\rho: vE \to \field F_2$ and $\sum_{e \in vE} \rho(e) = 1$ is stabilised (an 
\emph{odd} node).
These constraints are encoded by a simple graph gadget.
Although it seems that for each of these exponentially many choices we obtain a 
different CFI-graph, there really are, up to isomorphism, only two such graphs
which in turn are determined by the parity of the number of odd nodes.
Very roughly, the reason is that we can transpose, or \emph{twist}, two 
copies $e_0, e_1$ of each an edge $e$ and move this twist along a path (in the 
\emph{connected} graph $\mcG$) to iteratively resolve pairs of odd nodes.

In order to generalise this construction to $\field F_q$ we take for 
every edge $e \in E$ a \emph{directed cycle} of length $q$ over $q$ copies 
$e_0, e_1, \dots, e_{q-1}$ of the edge $e$.
We then add similar constraints for sets of incident edges as above, but 
naturally, instead of having only two different kinds of such constraints, we 
have one for each value $0, 1, \dots, q-1 \in \field F_q$.
Now, instead of twisting pairs of edges, we consider 
\emph{cyclic shifts} of length $\leq q$ on the edge classes $e_0, e_1, \dots, 
e_{q-1}$ which respect the cycle relation. Again, these shifts can be 
propagated along paths in the original graph $\mcG$ and, with a reasoning 
analogous to the original approach, it turns out that there are, up to 
isomorphism, only $q$ different types of generalised CFI-graphs over $\field 
F_q$. We remark that the same kind of generalisations has been studied, 
for example, in~\cite{Ho10,To04}.

\newcommand{\CFI}{\ensuremath{\logic{CFI}_q}}

\smallskip
Let us formalise the above described intuitions.
We start with an \emph{(undirected), connected} 
and \emph{ordered} graph $\mcG = (V,\leq, E)$.
Let $C, I$ and $R$ be binary relation symbols.
We set $\tau \defeq \{ \preceq, C, I, R \}$.
We define for every prime $q$ and every sequence of \emph{gadget values} 
$\vec d = (d_v)_{v \in V} \in \inseg q^V$ a $\tau$-structure $\CFI(\mcG, 
\vec d)$ which we call a \emph{CFI-structure over 
$\mcG$}.
For the following construction we agree that arithmetic is modulo $q$ so that 
we can drop the operator ``$\text{mod } q$'' in 
statements 
of the form $x 
= y \modulo q$ and $x + y \,\,\modulo q$ for the sake of better readability. 
For what follows, let $E(v) \subseteq E$ denote the set of \emph{directed} 
edges starting in $v$. 
Since $\mcG$ is an undirected graph, this means that for each undirected edge 
$\{v,w\}$ of $\mcG$ we have $(v,w) \in E(v)$ and $(w,v) \in E(w)$.
The construction is illustrated in Figure~\ref{fig:cfiq}.

\begin{itemize}
 \item The \emph{universe} of $\CFI(\mcG, \vec d)$ consists of \emph{edge 
nodes} and \emph{equation nodes}.
\begin{itemize}
\item The set of \emph{edge nodes} $\hat{E}$ is defined as $\hat E \defeq
\bigcup_{e \in E} \hat{e}$ where
 for every \emph{directed} edge $e \in E$ we let
 the \emph{edge class} $\hat{e} = \{ e_0, e_1, 
\dots, e_{q-1}\}$ consist of $q$ distinct copys of $e$. 
In particular, for every edge $e = (v,w) \in E$ and its reversed 
edge $e^{-1} \defeq f = (w,v) \in E$ the sets $\hat{e}$ and $\hat{f}$ are 
disjoint. 
We say that two such edges (or edge classes) are \emph{related}. 
\item The set of \emph{equation nodes} $\hat{V}$ is defined as $\hat V \defeq 
\bigcup_{v \in V} \hat{v}^{\vct d(v)}$ where for every 
vertex $v \in V$ and $d \in \inseg q$  the \emph{equation class} 
$\hat{v}^{d}$ 
consist of all functions $\rho: E(v) \to \inseg q$ which satisfy $\sum 
\rho \defeq \sum_{e \in E(v)} \rho(e) = d$.
\end{itemize}

 \item The \emph{linear preorder} $\preceq$ orders the edge 
classes according to the linear order induced by $\leq$ on $E$. More 
precisely, we let $\hat e \preceq \hat f$ whenever $e \leq f$.
Similarly, $\preceq$ orders the equation classes according to the 
order of $\leq$ on $V$, i.e.\ $\hat v \preceq \hat w$ if $v \leq w$. Moreover, 
we let $\hat e \preceq \hat v$ for edge classes $\hat e$ and  
equation classes $\hat v$.

 \item The \emph{cycle relation} $C$ contains a directed 
cycle of length $q$ on each of the edge classes $\hat{e}$ for $e \in 
E$, i.e.\ $C = 
\{ (e_i, e_{i+1}) : i \in \inseg q,  e\in E \}$.
 \item The \emph{inverse relation} $I$ connects two related edge classes 
by pairing additive inverses. More precisely, let 
$e=(v,w)\in E$ and $f = (w,v)\in E$. Then $I$ contains all 
edges $(e_x, f_y)$ with $x + y = 0$ for $x, y \in \inseg q$.

 \item The \emph{gadget relation} $R$  is defined as $R \defeq \bigcup_{v 
\in V} R^{\vct d(v)}_v$ where for $v \in V$ and $d \in \inseg q$ the relation 
$R^{d}_v$ is given as
\[ R^{d}_v \defeq \{ (\rho, e_{\rho(e)}) : \rho \in \hat{v}^{d}, e \in 
E(v) \}.\]
\end{itemize}

\begin{figure}[ht]
 \centering
\begin{tikzpicture}[>=stealth]
 
 %
 %
\coordinate (vertexv) at (-4.3,-1.8);

\draw[draw=lightgray, fill=white, thick] ($(vertexv) + 
(0,-0.2)$) 
ellipse (6em and 9em);

\node[font=\Large] at ($(vertexv) + (0,-2.7)$) {$\mcG$};
\draw [>=open triangle 90,->,very thick,lightgray ] ($(vertexv) + 
(1.67,1.7)$) arc (130:30:1.5cm);

\node[draw,fill,circle, inner sep=0.8pt, label={[label 
distance=0.4em]20:{\Large $v$}}] (v) at (vertexv)  {};
\node[draw,fill,circle, below left of=v, node distance =6em, inner 
sep=0.8pt, 
label=below:{\Large $a$}] (a)  {};
\node[draw,fill,circle, above of = v, node distance =6em, inner sep=0.8pt, 
label=above:{\Large $b$}] (b)  {};
\node[draw,fill,circle, below right of=v, node distance =6em, inner 
sep=0.8pt, 
label=below:{\Large $c$}] (c)  {};
\draw[->,thin] (v) to [bend left=20] node[below right] {$e^a$} (a) ;
\draw[->,thin] (a) to [bend left=20] node[above left] {$f^a$} (v) ;
\draw[->,thin] (v) to [bend left=20] node[left] {$e^b$} (b) ;
\draw[->,thin] (b) to [bend left=20] node[right] {$f^b$} (v) ;
\draw[->,thin] (v) to [bend left=20] node[above right] {$e^c$} (c) ;
\draw[->,thin] (c) to [bend left=20] node[below left] {$f^c$} (v) ;




\begin{scope}
\node (posb) at (1.8,-0.5) {};
\node (posbrecx) at ($(posb) + (-0.7,-0.5)$) {};
\node (posbrecy) at ($(posb) + (2,0.6)$) {};

\pgfmathsetmacro{\hoehe}{1.13} 
\pgfmathsetmacro{\breite}{3.11}
\coordinate (rbe1) at ($(posb) + (-0.85,-0.55)$);
\coordinate (rbe2) at ($(rbe1) + (0, \hoehe)$);
\coordinate (rbe3) at ($(rbe2) + (\breite, 0)$);
\coordinate (rbe4) at ($(rbe3) + (0, - \hoehe)$);
\draw[draw=black, fill=lightgray, fill opacity=0.3, thin] 
(rbe1) -- (rbe2) -- (rbe3) -- (rbe4) --cycle;

 \node[draw,fill,circle, inner sep=0.5pt, 
 label={[label distance=0.2em]180:{\footnotesize $e^b_0$}}] (eb0) at 
(posb) 
{};
\node[draw,fill,circle, inner sep=0.5pt, 
label={[label distance=0em]120:{\footnotesize $e^b_1$}},right 
of=eb0,node 
distance=2em] (eb1)  {};
\node[draw,fill,circle, inner sep=0.5pt, 
label={[label distance=0.2em]0:{\footnotesize $e^b_2$}},right of=eb1,node 
distance=2em] (eb2)  {};
 
 \draw[->,thin] (eb0) -- (eb1);
 \draw[->,thin] (eb1) -- (eb2);
 \draw[->, thin] (eb2) to  [bend left=85] (eb0) ;
 \node [below of = eb1, node distance=0.8em, font=\footnotesize] {$C$} ;

 \node (posb2) at ($(posb) + (0,2)$) {};
\pgfmathsetmacro{\hoehe}{1.13} 
\pgfmathsetmacro{\breite}{3.11}
\coordinate (rbf1) at ($(rbe1) + (0,2)$);
\coordinate (rbf2) at ($(rbf1) + (0, \hoehe)$);
\coordinate (rbf3) at ($(rbf2) + (\breite, 0)$);
\coordinate (rbf4) at ($(rbf3) + (0, - \hoehe)$);
\draw[draw=black, fill=hellblau, fill opacity=0.3, thin] 
(rbf1) -- (rbf2) -- (rbf3) -- (rbf4) --cycle;
 
 \node[draw,fill,circle, inner sep=0.5pt, 
 label={[label distance=0.2em]180:{\footnotesize $f^b_0$}}] (fb0) at 
(posb2) {};
\node[draw,fill,circle, inner sep=0.5pt, 
label={[label distance=0em]240:{\footnotesize $f^b_1$}},right 
of=fb0,node 
distance=2em] (fb1)  {};
\node[draw,fill,circle, inner sep=0.5pt, 
label={[label distance=0.2em]0:{\footnotesize $f^b_2$}},right of=fb1,node 
distance=2em] (fb2)  {};
 
 \draw[->,thin] (fb0) -- (fb1);
 \draw[->,thin] (fb1) -- (fb2);
 \draw[->, thin] (fb2) to [bend right=85] 
 node[below left,font=\footnotesize] {$C$} (fb0) ;
 
 \draw[-,thin] (fb0) to (eb0) ;
 \draw[-,thin] (fb2) to (eb1) ;
 \draw[-,thin] (fb1) to (eb2) ;
 \node (posbI) at ($(eb2)!0.5!(fb2)$) {$I$};
\end{scope}




\begin{scope}
 
\node (posa) at (1,-4.5) {};

\pgfmathsetmacro{\hoehe}{0.8} 
\pgfmathsetmacro{\breite}{2.2}
\coordinate (rae1) at ($(posa) + (-1.9,1.25)$);
\coordinate (rae2) at ($(rae1) + (\hoehe, \hoehe)$);
\coordinate (rae3) at ($(rae2) + (\breite, - \breite)$);
\coordinate (rae4) at ($(rae3) + (- \hoehe, - \hoehe)$);
\draw[draw=black, fill=lightgray, fill opacity=0.3, thin] 
(rae1) -- (rae2) -- (rae3) -- (rae4) --cycle;

 \node[draw,fill,circle, inner sep=0.5pt, 
 label={[label distance=0.05em]340:{\footnotesize $e^a_0$}}] (ea0) at 
(posa) 
{};
\node[draw,fill,circle, inner sep=0.5pt, 
label={[label distance=0.1em]270:{\footnotesize $e^a_1$}},above left
of=ea0,node 
distance=2em] (ea1)  {};
\node[draw,fill,circle, inner sep=0.5pt, 
label={[label distance=0.05em]120:{\footnotesize $e^a_2$}},above left 
of=ea1,node 
distance=2em] (ea2)  {};
 
 \draw[->,thin] (ea0) -- (ea1);
 \draw[->,thin] (ea1) -- (ea2);
 \draw[->, thin] (ea2) to [bend left=85] 
 node[left,font=\footnotesize] {$C$} (ea0) ;

 \node (posa2) at ($(posa) + (-1.4,-1.4)$) {};
\pgfmathsetmacro{\hoehe}{0.8} 
\pgfmathsetmacro{\breite}{2.2}
\coordinate (raf1) at ($(rae1) + (-1.5,-1.5)$);
\coordinate (raf2) at ($(raf1) + (\hoehe, \hoehe)$);
\coordinate (raf3) at ($(raf2) + (\breite, - \breite)$);
\coordinate (raf4) at ($(raf3) + (- \hoehe, - \hoehe)$);
\draw[draw=black, fill=hellblau, fill opacity=0.3, thin] 
(raf1) -- (raf2) -- (raf3) -- (raf4) --cycle;

 \node[draw,fill,circle, inner sep=0.5pt, 
 label={[label distance=0.05em]340:{\footnotesize $f^a_0$}}] (fa0) at 
(posa2) {};
\node[draw,fill,circle, inner sep=0.5pt, 
label={[label distance=0.05em]0:{\footnotesize $f^a_1$}},above left 
of=fa0,node 
distance=2em] (fa1)  {};
\node[draw,fill,circle, inner sep=0.5pt, 
label={[label distance=0.05em]120:{\footnotesize $f^a_2$}},above left 
of=fa1,node 
distance=2em] (fa2)  {};
 
 \draw[->,thin] (fa0) -- (fa1);
 \draw[->,thin] (fa1) -- (fa2);
 \draw[->, thin] (fa2) to [bend right=85] 
 node[right,font=\footnotesize] {$C$} (fa0) ;
 
 \draw[-,thin] (fa0) to (ea0) ;
 \draw[-,thin] (fa2) to (ea1) ;
 \draw[-,thin] (fa1) to (ea2) ;
 \node (posaI) at ($(ea2)!0.5!(fa2)$) {$I$};
 \end{scope}




\begin{scope}
 
\node (posc) at (4,-4.5) {};

\pgfmathsetmacro{\hoehe}{0.8} 
\pgfmathsetmacro{\breite}{2.2}
\coordinate (rce1) at ($(posc) + (-0.2,-0.95)$);
\coordinate (rce2) at ($(rce1) + (- \hoehe, \hoehe)$);
\coordinate (rce3) at ($(rce2) + (\breite, \breite)$);
\coordinate (rce4) at ($(rce3) + (\hoehe, - \hoehe)$);
\draw[draw=black, fill=lightgray, fill opacity=0.3, thin] 
(rce1) -- (rce2) -- (rce3) -- (rce4) --cycle;

 \node[draw,fill,circle, inner sep=0.5pt, 
 label={[label distance=0.05em]250:{\footnotesize $e^c_0$}}] (ec0) at 
(posc) {};
\node[draw,fill,circle, inner sep=0.5pt, 
label={[label distance=0.1em]270:{\footnotesize $e^c_1$}},above right
of=ec0,node 
distance=2em] (ec1)  {};
\node[draw,fill,circle, inner sep=0.5pt, 
label={[label distance=0.05em]20:{\footnotesize $e^c_2$}},above right 
of=ec1,node 
distance=2em] (ec2)  {};
 
 \draw[->,thin] (ec0) to (ec1);
 \draw[->,thin] (ec1) to  (ec2);
 \draw[->, thin] (ec2) to [bend right=85] (ec0) ;
 \node[font=\footnotesize] at ($(ec1)+(-0.35,0.05)$)  {$C$};

 \node (posc2) at ($(posc) + (1.4,-1.4)$) {};
\pgfmathsetmacro{\hoehe}{0.8} 
\pgfmathsetmacro{\breite}{2.2}
\coordinate (rcf1) at ($(rce1) + (1.5,-1.5)$);
\coordinate (rcf2) at ($(rcf1) + (-\hoehe, \hoehe)$);
\coordinate (rcf3) at ($(rcf2) + (\breite,  \breite)$);
\coordinate (rcf4) at ($(rcf3) + (\hoehe, - \hoehe)$);
\draw[draw=black, fill=hellblau, fill opacity=0.3, thin] 
(rcf1) -- (rcf2) -- (rcf3) -- (rcf4) --cycle;

 \node[draw,fill,circle, inner sep=0.5pt, 
 label={[label distance=0.05em]250:{\footnotesize $f^c_0$}}] (fc0) at 
(posc2) {};
\node[draw,fill,circle, inner sep=0.5pt, 
label={[label distance=0.05em]180:{\footnotesize $f^c_1$}},above right 
of=fc0,node 
distance=2em] (fc1)  {};
\node[draw,fill,circle, inner sep=0.5pt, 
label={[label distance=0.05em]20:{\footnotesize $f^c_2$}},above right 
of=fc1,node 
distance=2em] (fc2)  {};
 
 \draw[->,thin] (fc0) -- (fc1);
 \draw[->,thin] (fc1) -- (fc2);
 \draw[->, thin] (fc2) to [bend left=85] 
 node[left,font=\footnotesize] {$C$} (fc0) ;
 
 \draw[-,thin] (fc0) to (ec0) ;
 \draw[-,thin] (fc2) to (ec1) ;
 \draw[-,thin] (fc1) to (ec2) ;
 \node (poscI) at ($(ec2)!0.5!(fc2)$) {$I$};
 \end{scope}




 \node[draw=gray,fill=hellblau,rectangle, inner sep=1.5pt, 
 label={[label distance=0.15em]110:{\footnotesize $\rho=(2,1,0)$}}] (in210) at 
($(eb0) + (-0.5,-1.8)$) {};

\draw[-,darkgray] (in210) to (ea2) ;
\draw[-,darkgray] (in210) to (eb1) ;
\draw[-,darkgray] (in210) to (ec0) ;

\node[draw=gray,fill=hellblau,rectangle, inner sep=1.5pt, 
 label={[label distance=0.15em]20:{\footnotesize $\rho=(1,1,1)$}}] (in111) at 
($(eb2) + (0.5,-1.8)$) {};
  
\draw[-,darkgray] (in111) to (ea1) ;
\draw[-,darkgray] (in111) to (eb1) ;
\draw[-,darkgray] (in111) to (ec1) ;

\node[font=\large,darkgray] (Rdots) at ($(in210)!0.5!(in111)$) {$\cdots$};
\node[above of=Rdots, font=\large, darkgray, node distance=1.5em] (Rdots) {$R$};

\end{tikzpicture}  
\caption{Generalised CFI-construction for the $v$-gadget where $q=3$ and $\vec 
d(v) = 0$}
\label{fig:cfiq}
\end{figure}
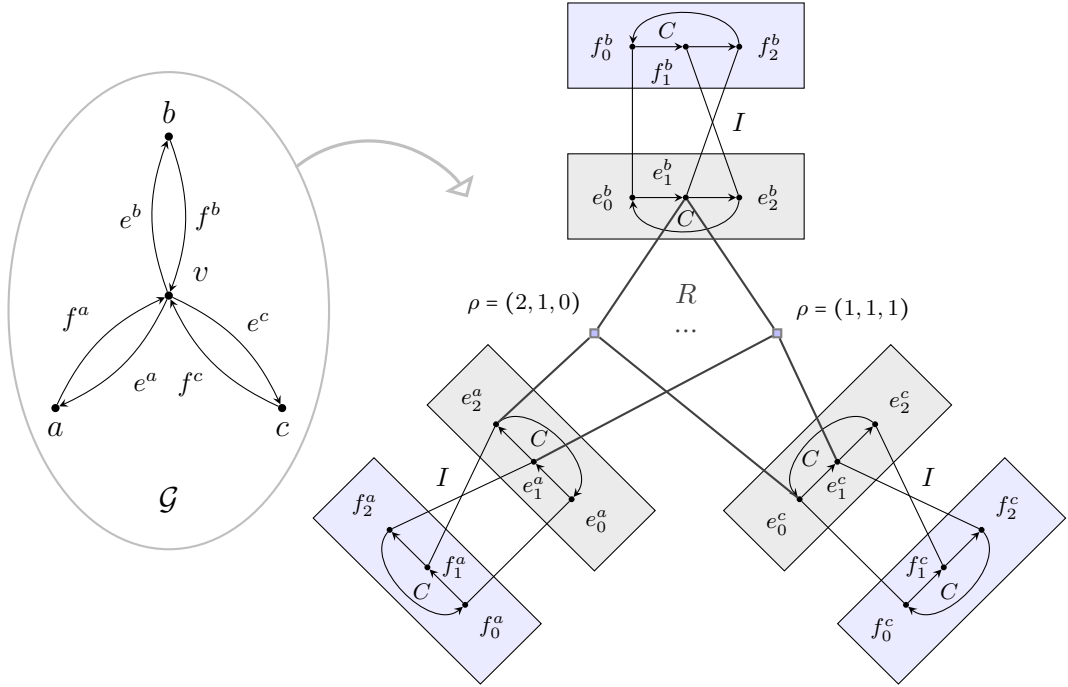

\newcommand{\gadget}{$\text{gadget}$}
At first glance our construction associates to every 
graph~$\mcG$ (with the above properties) and to each
sequence of gadget values $\vec d \in \inseg q^V$  a different structure 
$\CFI(\mcG, \vec d)$.
However, for each graph $\mcG$ with the above properties there really are, up 
to isomorphism, only $q$ different CFI-structures $\CFI(\mcG,\vct d)$.
In fact, the value $\sum \vec d \defeq \sum_{v \in V} \vec d(v)$ 
completely determines the isomorphism class of a CFI-structure over $\mcG$.

\medskip
To obtain this characterisation, we  analyse the automorphism group 
of CFI-structures 
and, more generally, the set of isomorphisms 
between two structures $\mfA = \CFI(\mcG, \vec d_1)$ and 
$\mfB = \CFI(\mcG, \vec d_2)$.
For such structures we know that the set $\hat E$ of edge nodes, the linear 
preorder $\preceq$ on $\hat E$, the cycle relation $C$ and the inverse relation 
$I$ do not depend on the sequence of gadget values. 
This means that each possible isomorphism $\pi$ which maps $\mfA$ to $\mfB$ 
induces an automorphism of the common substructure
$\mfC \defeq (\hat{E}, (\preceq \,\, \upharpoonleft \hat E), C, I)$ which only 
depends on $\mcG$ but not on $\vec d \in \inseg q^V$.
Thus  
\[ (\Iso(\mfA,\mfB) \upharpoonleft \hat E) \subseteq \Gamma \defeq 
\Aut(\mfC) \leq \Sym(\hat E).\]

Let $\pi \in \Gamma$.
The linear preorder $\preceq$ on $\hat{E}$ and the cycle 
relation $C$ enforce that $\pi$ is the composition of
 cyclic shifts on the individual edge classes $\hat{e}$, i.e.\ 
 $\pi \in \prod_{e \in E} 
\gengroup{ \pcycle {e_0 e_1 \cdots e_{q-1}} }  \leq 
\Sym(\hat E)$.
It is convenient to identify the group $\prod_{e \in E} \gengroup{ 
\pcycle {e_0 e_1 \cdots e_{q-1}} }$ with the vector space $\field F_q^E$ in the 
obvious way. 

In addition, the inverse relation $I$ enforces that cyclic shifts for 
pairs of related edge classes are inverse to each other in the following sense: 
let $e = (v,w) \in E$ and $f = (w,v) \in E$ be a pair of related edges.
Assume that we have a permutation $\pi \in \field F_q^E$ such that $\pi(e) = x$ 
and $\pi(f) = y$. We have $(e_0,f_0) \in I$. Hence, if $\pi$ is supposed 
to be an automorphism of $\mfC$ then we have $\pi(I) = I$ and thus 
$(e_x, e_y) \in I$ which means that $x + y = 0$.

In conclusion, it follows that $\Gamma \leq \field F_q^E$ is the subgroup of 
$\field F_q^E$ 
which contains all $E$-vectors $\pi \in \field F_q^E$ with
the property that $\pi(e) + \pi(f) = 0$ for pairs of related edges 
$e, f \in E$. 
Again we remind the reader that $\Gamma$ only depends on $\mcG$ but 
not on $\vec d \in \inseg q^V$. If we want to stress this dependence, then we 
sometimes write $\Gamma(\mcG)$ but usually we omit $\mcG$ if the graph is 
clear from the context.

Now, given a CFI-structure $\mfA = \CFI(\mcG,\vec d)$, we define for 
each vertex $v \in V$ the \emph{$v$-gadget} as the set $\gadget(v) \defeq \hat 
v^{d(v)} \uplus \bigcup_{e \in E(v)} \hat e$.

\begin{lemma}
Let $\mfA = \CFI(\mcG,\vec d)$ and let $\pi \in 
\Gamma$. 
Then there is precisely one extension $\hat \pi$ of $\pi$
to $\hat E \uplus \hat V$ such that $\hat \pi (\mfA)$ is a CFI-structure 
over $\mcG$.
\end{lemma}
\begin{proof}
 Let $\rho \in \hat v = \hat v^{\vct d(v)}$ for some $v\in V$. We show that 
under 
the 
assumption that $\hat \pi(\mfA)$ is a CFI-structure over $\mcG$ the 
action of $\pi$ on $\hat{E}$ determines $\hat \pi ( \rho )$.

We have that $(\rho, e_{\rho(e)}) \in R$ for all $e \in E(v)$. Hence for a 
potential isomorphism $\hat \pi$ we must have that 
$(\hat \pi (\rho), \pi( e_{\rho(e)} )) \in R'$ (for a gadget relation 
$R'$ of a CFI-structure over $\mcG$). 
Since we have $\pi( e_{\rho(e)}) = e_{\rho(e) + \pi(e)}$, it follows by the 
definition of CFI-structures that the function $\hat 
\pi(\rho) : E(v) \to \inseg q$ is determined as
$(\hat \pi(\rho))(e) = \rho(e) + \pi(e)$ which in turn only depends on the 
action of $\pi$ on the edge classes $\hat{e}$ for $e \in E(v)$.
\end{proof}

\smallskip
The preceding lemma shows that $\Iso(\mfA,\mfB)$ can be identified with 
a subset of $\Gamma$. In fact, the set 
$\Aut(\mfA)$ turns out to be a subgroup of $\Gamma$ of which $\Iso(\mfA,\mfB)$ 
is a coset in $\Gamma$.
Specifically, we saw that every $\pi \in \Gamma$ can uniquely be 
identified with an isomorphism of CFI-structures over $\mcG$ by setting
$\pi(\rho) = \rho + \pi$ for $\rho \in \hat v^{d}$ . As a consequence, 
this 
means that $\pi(\hat v^d) = \hat v^{d_*}$ where $d_* = d + \sum_{e \in 
E(v)} \pi(e)$ and that 
\[ \pi(R_v^{d}) = \{ (\rho + \pi, e_{\rho(e) + \pi(e)} ) : (\rho, 
e_{\rho(e)}) \in R_v^d\} = R^{d_*}_v.\]
In particular, $\pi$ stabilises the relation $R_v^{d}$ if, 
and only if, $\sum_{e \in E(v)} \pi(e) = 0$.

\begin{lemma}\label{lemma:action:on:cfi}
$\Gamma$ acts on $\{ \CFI(\mcG,\vec d) : \vec d \in \inseg q^V  
\}$. For $\pi \in \Gamma$ we have  
\[ \pi( \CFI(\mcG,\vec d) ) = \CFI(\mcG,\vec d_*) \text{ where } 
\vec d_*(v) = (\vec d(v) + \sum_{e \in E(v)}  \pi(e)).\]
\end{lemma}

\begin{lemma}\label{lemma:gencfi:isoclasses}
 Let $\vec d, \vec d_* \in (\inseg q)^V$ be two sequences of gadget values.
  Then $\CFI(\mcG, \vec d) \isom \CFI(\mcG, \vec d_*)$ if, and only if, $\sum 
\vec 
d= \sum \vec d_*$.
\end{lemma}
\begin{proof}
Let $\pi \in \Gamma$ such that 
$\pi(\CFI(\mcG, \vec d)) = \CFI(\mcG, \vec d_*)$.
By Lemma~\ref{lemma:action:on:cfi} this means that 
$\vec d_*(v) = (\vec d(v) + \sum_{e \in E(v)}  \pi(e))$ for $v \in V$.
Thus $\sum_{v \in V} \vec d_*(v) = \sum_{v \in V} \vec d(v) + \sum_{v \in 
V} \sum_{e \in E(v)}  \pi(e) = \sum_{v \in V} \vec d(v) + \sum_{e \in E} 
\pi(e)$. Since for all pairs of related edges $e, f \in E$ we have $\pi(e) 
+ \pi(f) = 0$ the claim follows.

\smallskip
For the other direction we proceed by induction on the number $i$ of 
vertices $v \in V$ such that $\vec d (v) \neq \vec d_*(v)$.
If no such vertex exists, then the claim is trivial.
Otherwise, because of our assumption, there exist at least two such
vertices $v,w \in V$, $v \neq w$.
Since $\mcG$ is connected we find a simple path 
\[ \bar p: v = v_0 \stackrel{E}{\longrightarrow} v_1  
\stackrel{E}{\longrightarrow} v_2 \stackrel{E}{\longrightarrow} \cdots 
\stackrel{E}{\longrightarrow} v_{m} = w\]
from $v$ to $w$ of length $m \geq 1$.
Consider the following $E$-vector $\pi \in \field F_q^E$ which is defined for 
$z \defeq \vec d_*(v) - \vec d(v)$ as
\[
 \pi(e) \defeq \begin{cases}
          z , & \text{ if } e = (v_i, v_{i+1}), 0 \leq i < m \\
          -z, & \text{ if } e = (v_{i+1}, v_{i}), 0 \leq i < m\\         
          0, &\text{ else.}  
          \end{cases}
\]
By the definition of $\pi$ it follows that $\pi \in \Gamma$.
Let $\pi( \CFI(\mcG, \vec d) ) = \CFI(\mcG,\vec d_+)$. We claim that the 
number of $v \in V$ such that $\vec d_+(v) \neq \vec d_*(v)$ is at 
most $i-1$.
From Lemma~\ref{lemma:action:on:cfi} we know that 
$\vec d_+(v) = \vec d(v) + \sum_{e \in E(v)} \pi (e)$. For $v \in 
V$ it follows that
\begin{itemize}
 \item if $v \nin \{ v_0, \dots, v_m \}$, then $\vec d_+(v) = \vec 
d(v)$, and
 \item if $v = v_0$, then $\vec d_+(v) = \vec d(v) + z = \vec d_*(v)$, and
 \item if $v=v_j$ for $1 \leq j < m$, then
 \[\vec d_+(v) = \vec d(v) + \pi(v_{j},v_{j-1}) + \pi(v_j,v_{j+1}) = \vec 
d(v) - z + z = \vec d(v), \text{ and }\]
 \item if $v = v_m$, then $\vec d_+(v) = \vec d(v) - z$.
\end{itemize}
Thus the claim follows from the induction hypothesis.
\end{proof}

The kind of isomorphism that we constructed in the proof of 
Lemma~\ref{lemma:gencfi:isoclasses} plays an important role
later on. Thus, for a simple path $\bp$ from $v_0$ to $v_m$ ($m \geq 1$) 
\[ \bar p: v = v_0 \stackrel{E}{\longrightarrow} v_1  
\stackrel{E}{\longrightarrow} v_2 \stackrel{E}{\longrightarrow} \cdots 
\stackrel{E}{\longrightarrow} v_{m} = w\]
as above and a constant $z \in \field F_q$ we denote this isomorphism by 
$\pi[{\bp},z] \in \Gamma$. 
In other words, if  we let $\sigma^z[e] \in \Gamma$ for $e\in E$ and $z \in 
\field F_q$ denote the $E$-vector which is defined as
\[ \sigma^z[e](f) = \begin{cases}
                    z,& \text{ if } f = e,\\
                    -z,& \text{ if } f = e^{-1},\\
                    0,&\text{ else,}
                   \end{cases} \]
then $\pi[{\bp},z] = \sigma^z[{(v_0,v_1)}] + \sigma^z[{(v_1,v_2)}] + \cdots + 
\sigma^z[{(v_{m-1},v_m)}]$.
Intuitively, the isomorphism $\pi[{\bp},z]$ 
allows us to simultaneously increase the gadget value at $v_0$ by $z$ and 
to decrease the gadget value at $v_m$ by $z$ while the induced twists are 
moved along the path $\bp$ through the gadget relations of the 
vertices $v_j$, $1 \leq j < m$, whose gadget value does not change.
A very important special case arises when $\bp$ is a simple cycle of length $m 
\geq 3$
\[ \bar p: v = v_0 \stackrel{E}{\longrightarrow} v_1  
\stackrel{E}{\longrightarrow} v_2 \stackrel{E}{\longrightarrow} \cdots 
\stackrel{E}{\longrightarrow} v_{m} = v.\]
Then for all values $z \in \field F_q$ the isomorphism $\pi[\bp,z] \in 
\Gamma$ is an \emph{automorphism} of CFI-structures over $\mcG$. 
We are going to use these automorphisms to show that it is possible to define 
in 
\FPC an ordering on the orbits of $\ell$-tuples as required 
by property~(\ref{item:property:classk:deforbits}).
It turns out that it therefore suffices to ensure that the graph $\mcG$ is 
sufficiently connected. 

\medskip
Recall that a graph $\mcG$ is \emph{$k$-connected}, for $k \geq 1$, 
if $\mcG$ contains more than $k$ vertices and if $\mcG$ stays connected 
when we remove any set of at most $k$ vertices. The \emph{connectivity} 
$\con(\mcG)$ 
of a graph $\mcG$ is the maximal $k \geq 1$ such that $\mcG$ is 
$k$-connected.
Moreover, the \emph{connectivity} $\con(\mfG)$ of a class $\mfG$ of 
graphs is the function $\con(\mfG): \N \to \N$ defined by
\[ n \mapsto \min_{\mcG \in \mfG, \card \mcG = n} \con(\mcG).\]
We are prepared to define the class $\mcK$: 
let $\mfG$ be a class of \emph{undirected, ordered} graphs
such that $\con(\mfG) \in \omega(1)$. 
Then we set
\[ \mcK = \mcK_q \defeq \{ \CFI(\mcG,\vec d) \, : \, \mcG=(V, \leq, E) \in 
\mfG, \vec d \in \inseg q^V\}.\]

\paragraph*{Verifying the required properties}
We proceed to show that $\mcK$ 
satisfies the required properties
(\ref{item:property:classk:autogroups}) - 
(\ref{item:property:classk:ckeqv}).

First of all, we saw that the automorphism group of each CFI-structure 
$\CFI(\mcG, \vct d)$ is a $\field F_q$-vector space, so 
property~(\ref{item:property:classk:autogroups}) clearly holds for the class 
$\mcK$.

The proof that $\mcK$ satisfies property~(\ref{item:property:classk:deforbits}) 
is more involved.
Let us fix the length $\ell \geq 1$ of tuples on which we want to 
define a linear preorder which identifies $\Delta_\mfA$-orbits.
By the choice of $\mcK$ it suffices to consider CFI-structures $\mfA = 
\CFI(\mcG, \vec d)$ over graphs $\mcG = (V, \leq, E)$ 
with $\con(\mcG) > (\ell + 2)$ since almost all structures in $\mcK$
satisfy this condition.
As above let $\Gamma \leq \field F^E_q$ denote the group that acts on the set 
of 
CFI-structures over $\mcG$ and let $A \defeq (\hat V \uplus \hat E)$ denote 
the universe of the 
CFI-structure $\mfA$.

\begin{definition}
 Let $\lambda \leq \ell$ and let $\ba \in A^\lambda$.

 \begin{enumerate}[(i)]
  \item Let $v \in V$. 
  We say that the vertex $v$ is \emph{marked (given the parameters $\ba$)} 
if for some $x \in \{ a_1, \dots, a_\lambda \}$ we have $x \in \hat v \, (= 
\hat{v}^{\vct d(v)})$.
  \item Let $e = (v,w) \in E$. 
  We say that the edge $e$ is \emph{marked (given the parameters~$\ba$)} 
if one of the vertices $v$ or $w$ is marked or if for some $x \in \{ a_1, 
\dots, a_\lambda \}$ we have that $x \in \hat e \cup \hat f$ where $f = 
(w,v) \in E$ is the edge related with $e$.
 \end{enumerate}
 \end{definition}
 
\begin{lemma}\label{lemma:marked:identified}
  Let $\lambda \leq \ell$ and let $\ba \in A^\lambda$.
\begin{enumerate}[(a)]
 \item If $v \in V$ is marked, then the $v$-gadget can be 
identified in $\INFCkx{\ell +2}$ (using the parameters $\ba$), i.e.\ 
 for every $c \in \gadget(v)$  there exists a formula 
 $\vartheta(\bx,y) \in \INFCkx{\ell +2}$ 
 such that $\vartheta^\mfA(\ba) = \{ c \}$.
 \item If an edge $e \in E$ is marked, then the edge classes $\hat e$ and 
$\hat f$ for $f = e^{-1}$ are 
identified in $\INFCkx{\ell +2}$ (given the parameters $\ba$), i.e.\ 
 for every $c \in \hat{e} \uplus \hat{f}$  there exists a formula 
 $\vartheta(\bx,y) \in \INFCkx{\ell +2}$ 
 such that $\vartheta^\mfA(\ba) = \{ c \}$.
\end{enumerate}
\end{lemma}
\begin{proof}
  First of all, it is straightforward (even without using the parameters) 
to fix the $\preceq$-class of any element $c \in A$ in $\INFCkx{\ell + 2}$.
  Secondly, observe that if an element $\rho \in \hat v$ is fixed, then 
we can fix an element in each of the edge classes $\hat e$ for $e \in 
E(v)$ since $\rho$ is $R$-connected to 
precisely one vertex in each of these classes. 
Moreover, if we have fixed an element $x \in \hat e$ in some edge class 
$\hat e$, then we can simply use the cycle relation $C$ to identify each 
element $c \in \hat e$ via its $C$-distance to $a$ in $\INFCkx{\ell + 
2}$.
Finally, the inverse relation $I$ yields a definable bijection between 
related edge classes. 
\end{proof}

\begin{lemma}\label{lemma:markedstable}
 Let $\lambda \leq \ell$, $\ba \in A^\lambda$ and let $v \in V$ be a 
vertex that is not marked.
 Then for all edges $e, e' \in E(v)$, $e \neq e'$ which are not marked 
there exists $\pi \in \Aut(\mfA,\ba)$ such that $\pi(e) = -\pi(e')\neq 0$ and 
such that $\pi(f)=0$ for all $f \in E(v) \setminus \{ e, e' \}$.
 \end{lemma}
\begin{proof}
Let $e=(v,w)$ and $e' = (v,w')$ as above.
Then the vertices $w$ and $w'$ are not marked.

Consider the graph $\mcG'$ that results from $\mcG$ by removing the vertex 
$v$ and each marked vertex $y \in V$.
Let $V' \subseteq V$ denote the vertex set and $E' \subseteq E$ 
the edge relation of the graph $\mcG'$.
Moreover, let  $M \defeq \{a_1, \dots, a_\lambda \} \cap 
(\bigcup_{e \in E} \hat e)$.
We observe that $\card {V} - \card {V'} \leq \lambda - \card M + 1$.

For every $x \in M$ there is an edge $f \in E$ such that 
$x \in \hat f$. For each such edge $f$ that is also contained in the 
subgraph $\mcG'$ we delete one of its endpoints but \emph{neither 
the vertex $w$ nor the vertex $w'$} and denote the resulting subgraph by 
$\mcG''$ with vertex set $V'' \subseteq V'$ and edge relation $E'' 
\subseteq E'$. 
It still might happen that there is a parameter $x \in M$ 
such that $x \in \hat f$ for $f \in E''$. However, this can only occur if 
$f$ connects $w'$ and $w$. 
Since we removed at most $(\card V - \card V') + \card M \leq \lambda + 1 \leq 
(\ell +1)$ vertices from the graph $\mcG$ to obtain  $\mcG''$ 
and 
since $\con(\mcG) > (\ell + 2)$ we know that there is a simple path 
of length $m \geq 2$ (i.e.\ the path does not consist of a 
single edge between $w$ 
and $w'$) which connects $w$ and $w'$ in $\mcG''$:
\[ \bar p: w \stackrel{E''}{\longrightarrow} v_1  
\stackrel{E''}{\longrightarrow} v_2 \stackrel{E''}{\longrightarrow} \cdots 
\stackrel{E''}{\longrightarrow} v_{m-1} \stackrel{E''}{\longrightarrow} w'.\]

\noindent
We extend $\bp$ to a simple cycle $\bp_c$ in $\mcG$ from $v$ to 
$v$ by 
using the edges $(v,w), (v,w') \in E$:
\[ \bar p_c: v \stackrel{E}{\longrightarrow} w 
\stackrel{E}{\longrightarrow} v_1 
\stackrel{E}{\longrightarrow} v_2 \stackrel{E}{\longrightarrow} \cdots 
\stackrel{E}{\longrightarrow} v_{m-1} \stackrel{E}{\longrightarrow} w' 
\stackrel{E}{\longrightarrow} v.\]
Let $0 \neq z \in \inseg q$. We claim that $\pi \defeq \pi[\bp_c,z]$ satisfies 
the desired properties.

By the definition of $\pi$ it holds that $\pi(e) = z = - \pi(e')$. 
Let $x \in \{ a_1, \dots, a_\lambda\}$. Then we have $x \nin 
\bigcup_{i=1}^{m-1} \hat v_i \cup \hat w \cup \hat w' \cup 
\hat v$, since none of the vertices $v$, $w$ and
$w'$ is marked and since we removed any other marked vertex $y \in V$ 
from $\mcG$.

Moreover, for $f \in \{ (v,w), (w,v), (v,w'), (w',v) \}$ we have 
that $x \nin \hat f$ by our assumption that $e, e'$ are not marked.
Also for $f \in \{ (w, v_1), (w',v_{m-1}) \}$ we have $x \nin \hat f$ since 
otherwise we had removed the vertices $v_1$ and $v_{m-1}$ from $\mcG'$.
Finally, for
$f \in \bigcup_{i=1}^{m-2} \{ (v_i, v_{i+1}), (v_{i+1}, v_{i}) \}$ 
we have $x \nin \hat f$
since otherwise we had removed one of the endpoints of each such edge $f$ from 
$\mcG'$. Hence $\pi(x)=x$. 
Finally, since $v \nin V''$ we also have that $\pi(f) = f$ 
for all $f \nin E(v) \setminus \{e, e'\}$.
\end{proof}

\begin{lemma}\label{lemma:slvclass:orbit:ckeqv}
 Let $\lambda \leq \ell$ and let $\ba, \bb \in A^\lambda$.
 Then $(\mfA,\ba) \Ckeqvx{\ell+2} (\mfA,\bb)$ if, and only if, there exists 
$\pi \in \Aut(\mfA)$ such that $\pi(\ba) = \bb$.
\end{lemma}
\begin{proof}
We proceed by induction on the maximal position $1 \leq i \leq \lambda$ up to 
which the tuples $\ba$ and $\bb$ agree, i.e.\ such that for $1 \leq j < i$ we 
have $a_j = b_j$ and such that $a_i \neq b_i$.
Let $a \defeq a_i$ and $b \defeq b_i$. 
Then we have to show that there exists an automorphism 
$\pi \in \Aut (\mfA, a_1 \, \cdots \, a_{i-1} )$ such that 
$\pi(a) = b$.
Since $\ba$ and $\bb$ have the same $\INFCkx{\ell +2}$-type we know that 
$a$ and $b$ belong to the same $\preceq$-class. 
We choose $v \in V$ such that $a,b \in \,\gadget(v)$. 

In what follows, whenever we speak of \emph{marked vertices} or 
\emph{marked edges} then we implicitly refer to a marking with respect to 
the already fixed part of parameters $\{ a_1, \dots, a_{i-1} \}$.

Without loss of generality we may assume that the vertex $v$ is not 
marked (by an element $x \in \{ a_1, \dots, a_{i-1} \}$), because otherwise, by 
Lemma~\ref{lemma:marked:identified}, every element in $\gadget(v)$ can 
uniquely be identified in $\INFCkx{\ell+2}$.
We distinguish between the two cases where $a$ and $b$ are equation nodes 
and where $a$ and $b$ are edge nodes.

For the first case let $a, b \in \hat v$.
There exists a unique $\pi \in \field F_q^{E(v)}$ such that 
$\pi(a) = b$ and such that $\sum_{e \in E(v)} \pi(e) = 0$.
Moreover, this vector $\pi$ can easily be defined in 
$\INFCkx{\ell+2}$ given the elements $a$ and $b$. 
Now assume that one of the edges $e = (v,w) \in E(v)$ is marked but that 
$\pi(e)\neq 0$. 
Since the edge $e$ is marked, every element in $\hat{e}$ can uniquely be 
identified in $\INFCkx{\ell+2}$ 
by Lemma~\ref{lemma:marked:identified}.
However, since $a$ and $b$ are $R$-connected to \emph{different} elements 
in $\hat{e}$ (as $\pi(e) \neq 0$) this 
contradicts the fact that $\ba$ and $\bb$ have the same $\INFCkx{\ell 
+2}$-type.
Thus, for every edge $e \in E(v)$ we either have that $\pi(e) = 0$ or that $e$ 
is not marked. By induction on the number of edges $e \in E(v)$ with 
$\pi(e)\neq 0$ we show that $\pi$ can be extended to an 
automorphism in $\Aut(\mfA, a_1, \dots, a_{i-1})$.
Thus let us fix $e \in E(v)$ such that $\pi(e) \neq 0$. Since we have 
that $\sum_{f 
\in E(v)} \pi(f) = 0$ there has to be another edge $e' \in E(v)$ with $\pi(e') 
\neq 0$. We apply 
Lemma~\ref{lemma:markedstable}
to obtain an automorphism $\sigma \in \Aut(\mfA, a_1, \dots, a_{i-1})$ such 
that 
$\sigma(e) = \pi(e)$, $\sigma(e') = -\pi(e)$ and $\sigma(f) = 0$ for all $f \in 
E(v)\setminus \{ e, e'\}$.
Now consider $(\pi - \sigma) \in \field F_q^{E(v)}$. By the 
induction 
hypothesis we can extend this vector to an automorphism 
$\pi_* \in \Aut(\mfA, a_1, \dots, a_{i-1})$. But then $(\pi_* + \sigma) \in 
\Aut(\mfA, a_1, \dots, a_{i-1})$ is an extension of $\pi$.

For the second case assume that $a,b \in \hat{e}$ for some edge $e \in E(v)$.
As above we conclude that the edge $e$ is not marked.
Since $\con(\mcG) > (\ell +2)$ the minimal degree of each 
vertex in $\mcG$ is at least $(\ell + 4)$. 
Since the vertex $v$ is not marked there has to be 
another edge $e' \in E(v)$, $e \neq e'$ which is not marked.
Thus we can apply 
Lemma~\ref{lemma:markedstable}
to obtain an automorphism $\pi \in \Aut(\mfA, a_1, \dots, a_{i-1})$ 
such that $\pi(a) = b$ and $\pi(f) = 0$ for all $f \in E(v) \setminus \{ 
e, 
e' \}$. 
\end{proof}

\medskip
It is well-known that classes of $\INFCkx{\ell +2}$-equivalent tuples 
can be ordered in $\FPC$, see e.g.\ \cite{Ot97}.
Hence, it follows from our previous lemma that the class $\mcK$ satisfies 
property~(\ref{item:property:classk:deforbits}). 

\begin{lemma} \label{lemma:slv:classk:prop:12}
The class $\mcK$ satisfies the properties
(\ref{item:property:classk:autogroups}) and
(\ref{item:property:classk:deforbits}).
\end{lemma}

\medskip
Let us now turn our attention to 
property~(\ref{item:property:classk:ckeqv}).
In the next lemma we are going to show that for each $k \geq 1$ and each
sufficiently connected graph $\mcG \in \mfG$, the logic $\INFCk$ cannot 
distinguish between any pair of CFI-structures over $\mcG$ (although there 
exist non-isomorphic CFI-structures over $\mcG$).
\begin{lemma} \label{lemma:slv:classk:prop:4}
 Let $k \geq 1$ and let $\mcG=(V,\leq,E) \in \mfG$ such that $\con(\mcG) > 
k$.
 Then for all $\vec d, \vec d_* \in {\inseg q}^V$ it holds that
 \[ \CFI(\mcG, \vec d) \Ckeqv \CFI(\mcG, \vec d_*).\]
 Thus, the class $\mcK$ satisfies 
property~(\ref{item:property:classk:ckeqv}).
\end{lemma}
\begin{proof}
 Let $\mfA = \CFI(\mcG, \vec d)$ and let $\mfB = \CFI(\mcG, \vec d_*)$. 
 Without loss of generality we assume that $\mfA \nisom \mfB$.
 We show 
that Duplicator wins the $k$-pebble bijection game 
on $\mfA$ and $\mfB$. Let $z_a \defeq \sum_{v \in V} \vec d(v)$, let $z_b 
\defeq \sum_{v \in V} \vec d_*(v)$ and let $z \defeq z_b - z_a$.
As above, for $e=(v,w) \in E$ and $y \in \inseg q$ we let $\sigma^y[e] \in 
\Gamma = 
\Gamma(\mcG)$ denote the isomorphism which shifts the edge class $\hat e$ by 
$y$, the edge class $\hat f$ for $f = (w,v)$ by $-y$ and which stabilises all 
remaining classes, i.e.\
\[ \sigma^y[e](f) = \begin{cases}
                     z,&\text{ if } f = (v,w),\\
                     -z,&\text{ if } f = (w,v),\\
                     0,&\text{ else.}
                    \end{cases}
\]
Given a position $(\mfA, a_1, \dots, a_\ell, \mfB, b_1, \dots, b_\ell)$ in the 
$k$-pebble bijection game, we say that a pair $(v,\pi)$ with $v\in V$ and $\pi 
\in \Gamma(\mcG)$ is \emph{good} if:
\begin{itemize}
 \item the $v$-gadget is not marked (by the pebbled elements $a_1, \dots, 
a_\ell$ in $\mfA$ or, equivalently, by the pebbled elements $b_1, \dots, 
b_\ell$ in $\mfB$),
 \item $\pi(a_i) = b_i$ for $1 \leq i \leq \ell$, 
 \item $\pi(\mfA \setminus \hat v) = \mfB \setminus \hat v$, and
 \item $(\sigma^z[e]+ \pi)(\mfA \upharpoonleft \gadget(v)) = \mfB 
\upharpoonleft \gadget(v)$ for all $e \in E(v)$.
\end{itemize}
Intuitively this means that $\pi$ is nearly an isomorphism between $\mfA$ 
and $\mfB$ except for the gadget associated to vertex $v$.
Of course $\pi$ itself does not induce a bijection between the universes 
of the two CFI-structures (as otherwise $\mfA \isom \mfB$).
However, for each $e \in E(v)$ we can associate a bijection $\hat \pi_e: 
A \to B$ to $\pi$ which is defined as
\[ \hat \pi_e (x) = \begin{cases}
                   \pi(x) ,&\text{ if } x \nin \hat v,\\
                   (\sigma^z[e]+ \pi)(x) ,&\text{ if } x \in \hat v.
                  \end{cases}
\]

In what follows we show that Duplicator can play in such a way that 
after each round such a good pair $(v, \pi)$ exists. Obviously, if Duplicator 
can 
maintain this invariant this suffices for her to win the game.

Indeed we can find such a good pair $(v,\pi)$ 
by Lemma~\ref{lemma:gencfi:isoclasses} for the initial position $(\mfA, \mfB)$ 
of the game.
Let us now consider one round of the game which starts from a 
position $(\mfA, a_1, \dots, a_\ell, \mfB, b_1, \dots, b_\ell)$ for which a 
good 
pair $(v,\pi)$ exists.
First, Spoiler chooses a pair $i \leq k$ of pebbles which he removes from the 
game board (if the corresponding pebbles are placed at all). 
Duplicator then answers Spoiler's challenge by providing a bijection 
$\hat \pi_e$ for some edge $e \in E(v)$ which is not marked. 
Note that such an edge $e$ exists since $\con(\mcG) > k$ and thus each 
vertex has degree at least $k + 2$.
Spoiler picks a new pair $(a, \hat \pi_e(a)) \in A \times B$ of $\hat 
\pi_e$-related elements on which he places the $i$-th pair of pebbles.
By the properties of $\pi$ it immediately follows that the resulting mapping 
$\ba[i\mapsto a] \mapsto \bb[i \mapsto b]$ is a partial isomorphism. 
However, it might happen that Spoiler placed the $i$-th pair of pebbles on 
equation nodes $\hat v$ in the gadget associated to vertex $v$. In this case 
the 
pair $(v, \pi)$ is not good any longer.
So assume that Spoiler pebbled a new pair of elements $(a, \pi_e(a)) \in 
\hat v \times \hat v$.
Since the edge $e = (v,w)$ was not marked we know that $w$ is not 
marked.
Thus it is easy to see that the 
pair $(w, \sigma^z[e] +  \pi)$ is good. 
\end{proof}

To complete our proof we establish an $\FPSx q$-definable 
canonisation procedure on the class $\mcK$. 
The idea is as follows: given a CFI-structure $\mfA = \CFI(\mcG,\vec d)$ 
over a graph $\mcG$ and a value $z \in \inseg q$ we construct a 
linear equation system over $\field F_q$ which is solvable if, 
and only if, $\sum \vec d = z$. This linear equation system 
is $\FO$-definable in the structure $\mfA$ which shows that $\FPSx q$ can 
determine the isomorphism class of a CFI-structure over $\mcG$. Since the 
graph $\mcG$ is ordered it is easy to construct an ordered
representative from each isomorphism classes of CFI-structures over $\mcG$ 
which concludes our argument.

More specifically, let $\mcG = (V, \leq, E) \in \mfG$, let $\mfA = 
\CFI(\mcG, \vec d) \in \mcK$ and let $z \in \field F_q$. 
For our linear equation system we identify each element $e_i \in \hat{E}$ 
and each vertex $v \in V$ with a variable over $\field F_q$, i.e.\ we let 
$\mcV \defeq \hat{E} \uplus V$ be the set of variables.
The equations of the linear system are given as follows:

\eqcountreset
\begin{align}
 e_{i+1} &= e_i +1 \quad&\text{for all } e_i \in \hat E \eqcount
 \label{les:slvq:e1}\\
 e_{i} &= - f_{-i} \quad&\text{for related edges } e,f \in E 
\eqcount
\label{les:slvq:e2}
\\
 v &= \sum_{e \in E(v)} e_{\rho(e)} \quad&\text{for all } v \in V, \rho 
\in \hat v \eqcount
\label{les:slvq:e3}
\\
 z &= \sum_{v \in V} v.
 \label{les:slvq:e4}
 \eqcount
\end{align}

It is easy to see that this system is $\FO$-definable in $\mfA$. 
First of all, the equation (\ref{les:slvq:e4}) can be defined as a sum over the 
ordered set $V$. Moreover, we can 
express the equations of type (\ref{les:slvq:e1}) and (\ref{les:slvq:e2}) by 
using the cycle and inverse relation, respectively. 
Finally, the equations of type (\ref{les:slvq:e3}) can be expressed by using 
the gadget relation $R$.

\begin{lemma}
 The above defined system is solvable if, and only if, 
 $\sum \vec d = z$.
\end{lemma}
\begin{proof}
 If $\sum \vec d = z$ then it is easy to verify that we obtain a solution 
$\vec \sigma \in \field F_q^{\mcV}$ of the linear system by setting $\vec 
\sigma(e_i) = i$ and $\vec \sigma(v) = \vec d(v)$.
For the other direction, we show that a solution $\vec \sigma \in \field 
F_q^\mcV$ of 
this system defines an isomorphism $\pi$ between $\mfA$ and $\mfB = 
\CFI(\mcG,\vec d_+)$ where $\vec d_+(v) \defeq \vec \sigma (v)$. 
As a preparation, we let $\delta(e) \defeq \vec \sigma(e_{i}) - i$ for $e \in 
E$ and some $e_i \in \hat e$. Since $\vec \sigma$ is a solution, $\delta \in 
\field F_q^E$ is well-defined.
Now we obtain the isomorphism $\pi$ for $e_i \in \hat E$ and $\rho \in 
\hat V$ by setting
\begin{align*}
 \pi(e_i) &\mapsto e_{\sigma(e_i)} \\
 \pi(\rho) &\mapsto \rho + \delta.
\end{align*}
Using the equations (\ref{les:slvq:e1}) and (\ref{les:slvq:e2}) one 
easily verifies that $\pi$ respects the cycle relation $C$ and the inverse 
relation $I$.
Moreover, let $(\rho, e_{\rho(e)}) \in R$. Then 
\[ \pi(e_{\rho(e)}) = e_{\vec \sigma(e_{\rho(e)})}
\text{ and } \vec \sigma(e_{\rho(e)}) = \rho(e) + \delta(e).
\]
Thus, $\pi$ also respects $R$.
Finally, by the equations of type (\ref{les:slvq:e3}),  for all $v \in V$ and 
$\rho \in \hat v$ we have that
\[ \sum \rho + \delta = \sum_{e \in E(v)} \vec 
\sigma(e_{\rho(e)}) = \vec \sigma(v).\]
This shows that $\vec \sigma(v) = d_+(v)$ and that $\sum \vec d_+ = \sum_{v \in 
V} \vec \sigma (v) = z$ because of equation (\ref{les:slvq:e4}).
\end{proof}

\begin{lemma} \label{lemma:slv:classk:prop:3}
The class $\mcK$ satisfies the property
(\ref{item:property:classk:isopoly}).
\end{lemma}

This finishes our proof of Theorem~\ref{thm:fps:distinct:primes}.

\section{Solvability quantifiers vs.\ rank operators}
\label{sec:slv:rk}

In the previous section we obtained separation results for the 
extensions of $\FPC$ by solvability quantifiers (and rank operators) over 
different sets of primes.
One important step of our proof was to construct a class of structures on which 
the expressive power of the logics $\FPRx{\Omega}$ and $\FPSx{\Omega}$ 
coincides. 
Moreover, as we already mentioned in Section~\ref{sec:logics:linalgop}, most of 
the queries which are known to separate fixed-point logic with counting and 
rank logic can also be expressed in $\FPS$. 
This leads to the interesting question whether, in general, rank operators can 
be simulated by solvability quantifiers within fixed-point logic with counting.
In this context, it is worthwhile to remark that many other problems from 
linear algebra are known to sit in between of ``solving linear equation 
systems'' and ``computing the matrix rank'', for example, deciding whether two 
matrices are similar or equivalent, see~\cite{Pa10, Ho10, La11}.

\smallskip
In this section we solve a simplified version of this question and show that 
in the absence of fixed-points and, more importantly, in the 
absence of counting, rank operators are strictly more expressive than 
solvability quantifiers. The reader should note that rank 
operators can easily simulate counting terms but this does not hold 
for solvability quantifiers.

In order to state our main result formally, we first define for every
prime $p$ the extension $\FOSp$ of first-order logic (without counting) by 
solvability quantifiers over $\field F_p$. 
The crucial difference to the extension $\FORp$ of first-order logic by rank 
operators $\rkp$ is that the logic $\FOSp$ is a \emph{one-sorted} logic which 
does not have access to a counting sort.

\begin{definition}\label{def:fosp}
For every prime $p$, the logic $\FOSp$ results by extending the syntax of 
$\FO$ by the following
formula creation rule:
\begin{itemize}
 \item If $\phi(\bx,\by,\bz) \in \FOSp$, then
   $\psi(\bz) = (\slvp \, \bx, \by) \phi(\bx,\by, \bz)$ is an 
$\FOSp$-formula.
\end{itemize}
\noindent
The semantics of $\psi(\bz)$ are defined as above. 
For completeness, let $k =\card{\bx}$ and $\ell = \card{\by}$. A pair $(\mfA, 
\bz \mapsto \bc)$ with $\bc \in A^{\card z}$ defines an $I\times J$-matrix 
$M_\phi$ over $\{ 0, 1 \}
\subseteq \field F_p$ where $I = A^k$ and $J = A^\ell$ and where
$M_\phi(\ba,\bb) = 1$ if, and only if, $\mfA \models \phi(\ba,\bb,\bc)$.

Let $\onevec$ be the $I$-identity vector over $\field F_p$, i.e.\ 
$\onevec(\ba) = 1$ for all $\ba \in I$. Then $M_\phi$ and $\onevec$ determine
the linear
equation system $M_\phi \cdot \vct x = \onevec$ over $\field F_p$.
Now we let $\mfA \models \psi(\bc)$ if, and only if, $M_\phi 
\cdot \vct x = \onevec$ is solvable.
\end{definition}

Analogously to the definition of $\FPS$ in Section~\ref{sec:logics:linalgop}, 
the syntactic normal form of definable linear equation systems in the 
definition of $\slvp$-quantifier does not lead to a severe restriction (again, 
see Lemma~4.1 in~\cite{DaGrHoKoPa13}).

\medskip
Let us briefly summarise what is known about the logic $\FOSp$ (see
also \cite{DaGrHoKoPa13,Pa10}). First of all, it follows from \cite{DaGrHoLa09} 
that for
every prime $p$, the logic 
$\FOSp$ can express the symmetric transitive closure of definable 
relations. Hence, $\FOSp$ subsumes the logic $\STC$ and can express every 
\LOGSPACE-computable property of ordered structures. Secondly, it also follows
from \cite{DaGrHoLa09} that $\FOSx{2}$ can distinguish between the odd and even
version of a CFI-graph, which means that $\FOSx{2}$ cannot be a fragment of
$\FPC$. More generally, by adapting the CFI-construction for other fields
one can show that $\FOSx{p} \not\leq \FPC$ for all $p\in \bbP$ (see
e.g.\ \cite{Ho10}).

On the domain of ordered structures, the expressive power 
of $\FOSp$ can be characterised in terms of a natural complexity 
class: in \cite{BuHeDaMe91}, Buntrock~et.~al.\ introduced
the \emph{logarithmic space modulo counting classes} $\MODLx{k}$ for integers
$k\geq 2$. Analogously to the case of modulo counting classes for polynomial
time, the idea is to say that a problem is in $\MODLx{k}$ if there exists a
non-deterministic logspace Turing machine which verifies its inputs by producing
a number of accepting paths which is not congruent $0 \,\,\modulo k$. For the 
formal
definition we refer the reader to~\cite{BuHeDaMe91}. It turns out that, at least
for primes $p$, the class $\MODLx{p}$ is closed under many natural operations,
including all Boolean operations and even logspace Turing reductions
\cite{BuHeDaMe91,HeReVo00}. Furthermore, many problems from linear
algebra over $\field F_p$ are complete for $\MODLx{p}$. In particular this is
true for the solvability problem of linear equation systems over $\field F_p$
and for computing the matrix rank over $\field F_p$~\cite{BuHeDaMe91}.

Building on these insights, Dawar~et.~al.\ were able to show that for all
primes $p$, the logic $\FORx{p}$ captures $\MODLx{p}$ on the class of ordered
structures. It has been noted in~\cite{Pa10} that their proof shows that 
the same correspondence holds for the logic $\FOSx{p}$.
\begin{prop}[\cite{DaGrHoLa09},\cite{Pa10}]
 On ordered structures we have $ \FOSx{p} = \FORx{p} = \MODLx{p}$.
\end{prop}

Despite this characterisation over the class of ordered 
structures, the situation over general structures remained unclear. It easily 
follows that $\FOSx{p} \leq \FORx{p} \leq \FPRK$, but, 
so far, it 
has been open whether one, or even both, of these inclusions are strict.
In this section we are going to settle one of these questions:

\begin{theorem}
\label{thm:fos:for}
 For all primes $p$ we have $\FOSp < \FORp$ (over the class of sets 
$\Str(\emptyset)$).
\end{theorem}

In some sense, this result is not very surprising. 
Over the class of sets, the logic $\FORp$ captures the complexity class 
$\MODLx{p}$ since the size of a set is a complete invariant.
In contrast, the logic $\FOSp$ cannot access the counting sort 
and thus had to express properties over pure unordered sets which have 
the maximal amount of symmetries.
However, it is not obvious how one can turn this intuition into a formal 
argument. Strikingly, $\FOSp$ has non-trivial expressive 
power over sets. For instance, $\FOSp$ can determine the 
size of sets modulo $p$~\cite{Pa10}, and consequently, modulo $p^k$ for 
every fixed $k$ (since $n \equiv 0 \mod p^k$ if, and only if, 
$n \equiv 0 \mod p$ and ${n \choose p} \equiv 0 \mod p^{k-1}$). 
Note that fixed-point logic $\FP$, for example, collapses to 
first-order logic over sets.

In order to prove Theorem~\ref{thm:fos:for} we make use of the following 
strong normal form for \FOSp which has been established 
in Corollary~4.8 of~\cite{DaGrHoKoPa13}.
\begin{theorem}
\label{thm:normalform:fos}
Every formula $\vartheta(\bz) \in \FOSp$ is equivalent to an 
$\FOSp$-formula of
the form $(\slvp \, \bx_1,\bx_2) \alpha(\bx_1,\bx_2, \bz)$ where 
$\alpha(\bx_1,\bx_2,\bz)$ is quantifier-free.
\end{theorem}

\medskip
Similar to our approach in Section~\ref{sec:sep:fields}, the main idea 
for separating $\FOSp$ and $\FORp$ is to exploit the symmetries of definable 
linear equation systems.
More precisely, we are aiming at considerably reducing the size of an input 
linear equation system via an $\FORp$-definable transformation. For the 
remainder of this proof, let us fix a quantifier-free formula 
$\alpha(x_1, \dots, x_k, y_1, \dots, y_\ell)\in \FO(\emptyset)$ and a 
prime $p$. 
According to the semantics of $\FOSp$, the formula $\alpha$ defines in an
input structure $\mfA= (\inseg{n})$ of size $n$ the $\inseg{n}^k \times
\inseg{n}^\ell$-coefficient
matrix $M_n$ which is given for $\ba \in \inseg{n}^k, \bb \in
\inseg{n}^\ell$ as
\[ M_n(\ba, \bb) = \begin{cases}
                        1, &\text{if } \mfA \models \alpha(\ba, \bb) \\
                        0, &\text{otherwise.}    
                       \end{cases}  \]
Then $\mfA \models (\slvp \, \bx_1,\bx_2) \alpha(\bx_1,\bx_2)$ if the linear
equation system $M_n \cdot \vct x = \onevec$  over $\field F_p$ is solvable.
For convenience we set $I_n = \inseg{n}^k$ and $J_n = \inseg{n}^\ell$. 

Let $\Gamma = \Gamma_n = \Sym(\inseg{n})$. Then the group $\Gamma$ acts on $I_n$
and $J_n$ in the natural way. 
As in Section~\ref{sec:sep:fields} we identify the action of $\pi \in 
\Gamma$ with the multiplication by the associated $I_n \times 
I_n$-permutation matrix $\Pi_I$ and the $J_n \times J_n$-permutation 
matrix $\Pi_J$, respectively. Hence, for $\pi \in \Gamma$ we have 
\[ \Pi_I \cdot M_n \cdot \Pi_J^{-1} = M_n \quad\Leftrightarrow\quad 
\Pi_I \cdot M_n = M_n \cdot \Pi_J. \]

For what follows, we fix a prime $q$ which is distinct from $p$ 
and a subgroup $\Delta 
\leq \Gamma$ which is a $q$-group, i.e.\ $\card \Delta = q^m$ for some $m 
\geq 0$. The overall strategy is to use the $\Delta$-symmetries of the 
matrix $M_n$
to strongly reduce the size of the linear equation system $M_n \cdot \vec x =
\onevec$.
More precisely we claim that for $M_n^* \defeq \sum_{\pi \in \Delta} \Pi_I
\cdot M_n$ the
linear equation system $M_n \cdot \vec x = \onevec$ is solvable if, and only if,
$M_n^*\cdot \vec x = \onevec$ is solvable. First of all we note that for all
$\pi \in \Delta$ we have:
\begin{itemize}
 \item $\Pi_I \cdot M_n^* = \sum_{\lambda \in \Delta} \Pi_I \cdot \Lambda_I
\cdot M_n = \sum_{\pi \in \Delta} \Pi_I \cdot M_n = M_n^*$
 \item $M_n^* \cdot \Pi_J = 
 \sum_{\lambda \in \Delta} \Lambda_I \cdot M_n \cdot \Pi_J = 
 \sum_{\lambda \in \Delta} \Lambda_I \cdot \Pi_I \cdot M_n = M_n^*$.
\end{itemize}
To verify our original claim assume that $M_n^* \cdot \vec b =
\onevec$. Then we have
\[ \onevec = M_n^* \cdot \vec b = (\sum_{\pi \in \Delta} \Pi_I \cdot M_n)
\cdot \vec b = 
(\sum_{\pi \in \Delta} M_n \cdot \Pi_J) \cdot \vec b =
M_n \cdot \sum_{\pi \in \Delta} (\Pi_J \cdot \vec b). \]
For the other direction let $M_n \cdot \vec b = \onevec$. Then
$\sum_{\pi \in \Delta} \Pi_I \cdot M_n \cdot \vec b = \card{\Delta} \cdot
\onevec$, hence $(1 / \card{\Delta}) \cdot \vec b$ is a solution of the
linear equation system $M_n^* \cdot \vec x = \onevec$. Note that for this 
direction we require that $q$ and $p$ are co-prime as we have to divide by 
$\card \Delta$.

\smallskip
Since $M_n^*$ satisfies $\Pi_I \cdot M_n^* = M_n^* \cdot \Pi_J = M_n^*$ for
all $\pi \in \Delta$ we have 
\[ M_n^*(\ba,\bb) = M_n^*(\pi (\ba), \bb) = M_n^*(\ba,\pi(\bb)) \] for all $\ba
\in I_n, \bb \in J_n$ and $\pi \in \Delta$.
In other words, the entries of the $I_n\times J_n$-matrix $M_n^*$ are constant 
on the $\Delta$-orbits of the index sets $I_n$ and $J_n$. 
More specifically, if we let $I_n^\Delta$ and $J_n^\Delta$ denote the sets of
$\Delta$-orbits on $I_n$ and $J_n$, respectively, then $M_n^*$ can
be identified with the matrix $(M_n^* / \Delta)$ which is defined as 
\[ (M_n^* / \Delta): I_n^\Delta \times J_n^\Delta \to \field F_p,
([\ba],[\bb]) \mapsto M_n^*(\ba,\bb).\]
Note that, depending on the size of the group $\Delta$, the sets 
$I_n^\Delta$ and $J_n^\Delta$ can be noticeably smaller than the index 
sets $I_n$ and $J_n$. 
Hence our obvious strategy is to choose $\Delta$ as large as 
possible to obtain a much more compact linear equation system $M_n^* \cdot 
\vec x = \onevec$ which is  equivalent to the given one.

\medskip
Recall that the maximal $q$-subgroups $\Delta \leq \Gamma$  are 
 the \emph{$q$-Sylow groups} of $\Gamma$. 
It is well-known that for the case where $\Gamma=\Sym({\inseg n})$ these 
groups can be obtained via an inductive construction which we
want to explain here for the special case of $n$ being a power of $q$ (the 
general case can be handled similarly, see e.g.~\cite{Ha76}). Hence from 
now on, let us assume that $n = q^r$ for some $r \geq 1$.

First of all, we determine the size of $q$-Sylow groups of 
$\Gamma$. A simple induction shows that
the maximal $t \geq 1$ such that $q^t$  divides 
$n! = (q^r)!$
is given as
\[ t = q^{r-1} + q^{r-2} + \cdots + q + 1 = \frac{q^r-1}{q-1}.\]
In fact, we can write $(q^r)!$ as $(q^r)! = 1 \cdots (1 \cdot q) \cdots (2 \cdot
q) \cdots (q^{r-1} \cdot q)$.
Hence $t = t_* + q^{r-1}$ where $t_*$ is the maximal such that $q^{t_*}$
divides $(q^{r-1})!$.

In particular, if we denote for $n = q^r$ a $q$-Sylow of $\Sym(\inseg{n})$
by $\Delta_r$, then our argument from above shows that $\card{\Delta_1} =
q$ and that
\[ \card{\Delta_{r+1}} = \card{\Delta_r}^q \cdot q. \]

As it turns out, this equation already gives a hint about the algebraic
structure of $\Delta_r$. Indeed, $\Delta_{r+1}$ can be obtained as the 
\emph{wreath product} of 
$\Delta_{r}$ and the cyclic group $\field F_q$. Since $\Delta_1 = \field F_q$ 
it 
follows that $\Delta_r$ is the $r$-fold wreath product of the 
cyclic group $\field F_q$. We
decided to skip the
formal definition of the notion of wreath products and rather to directly
illustrate this concept for the particular case of the $q$-Sylow groups of 
$\Gamma = \Sym(\inseg n) = \Sym(\inseg {q^r})$.

To obtain an algebraic description of these groups, we inductively 
construct for $r \geq 1$ a $q$-Sylow subgroup $\Delta_r \leq \Sym(\inseg 
{q^r})$ together 
with a family of trees $\mcT_i^x$ for $i = 0, \dots, r$ 
and $x \in \inseg {q^{r-i}}$ such that the following properties
hold.
\begin{enumerate}[(I)]
 \item $\mcT_i^x$ is a complete $q$-ary tree of height $i$ whose leaves 
are labelled with elements from $\inseg n$.
 More precisely, the labels of the leaves of $\mcT_i^x$ form the set 
$\mcP_i^x = \{ x \cdot q^i, \dots, (x +1) \cdot q^i - 1 \}$ (note that 
$\mcP_i^x$ is the $x$-th block of the natural partition of $\inseg n$ into 
parts of size $q^i$).
\label{properties:sylow_tress:1}
 \item For all $i \leq r$ the group $\Delta_r$ transitively acts on the set $\{
\mcT^x_i : x \in \inseg{q^{r-i}} \}$ by applying
permutations $\delta \in \Delta_r$ to the labels of the leaves of the tress 
$\mcT_i^x$.
Moreover, for each $i \leq r$, the subgroup of $\Delta_r$ which point-wise 
stabilises the trees $\mcT_i^x$ is a normal subgroup of $\Delta_r$.
\label{properties:sylow_tress:2}
 \item We have $\Delta_1 \leq \Delta_2 \leq \cdots \leq \Delta_r$ where 
$\Delta_i$ acts on the set of labels $\mcP_i^0$ of the tree 
$\mcT_i^0$. More generally, for every block $\mcP_i^x$, the group $\Delta_r$
contains a subgroup $\Delta_r^{i,x} \leq \Delta_r$ which point-wise fixes
the elements of all blocks $\mcP_i^y$ for $y \neq x$ and whose action on
$\mcP_i^x$ corresponds to the action of $\Delta_i$ on $\mcP_i^0$.
\label{properties:sylow_tress:3}
\end{enumerate}

The inductive construction of the trees $\mcT_i^x$ is depicted in
Figure~\ref{fig:definition:trees_q_sylow}. 
To understand this construction better, it is quite
useful to think of elements $y \in \inseg n$ as being represented in their 
$q$-adic encoding, i.e.\
$y = y_0 + y_1 \cdot q + \cdots + y_{r-1} \cdot q^{r-1}$. 
Then we have that $y \in \mcP_r^0 = \inseg n$ and
\begin{itemize}
 \item $y \in \mcP_{r-1}^{y_{r-1}}$ 
 \item $y \in \mcP_{r-2}^{y_{r-2} +  y_{r-1} \cdot q}$
 \item $\dots$
 \item $y \in \mcP_{0}^{y_0 + \cdots + y_{r-1} \cdot q^{r-1}} = \mcP_0^y$.
\end{itemize}
Hence, the $q$-adic encoding of $y$ describes the unique path in the tree 
$\mcT_r^0$ from the root to the leaf $\mcT_0^y$.
The trees $\mcT_i^x$ clearly satisfy the properties stated
in~(\ref{properties:sylow_tress:1}).

For the inductive construction of the $q$-Sylow groups
$\Delta_r$ we first fix $\Delta_1$ as the cyclic group generated by the natural 
cyclic shift $\gamma = (0 \, 1 \, \cdots \, q-1)$ on the set $\mcP_1^0 = \{ 0, 
\dots, q-1\}$.

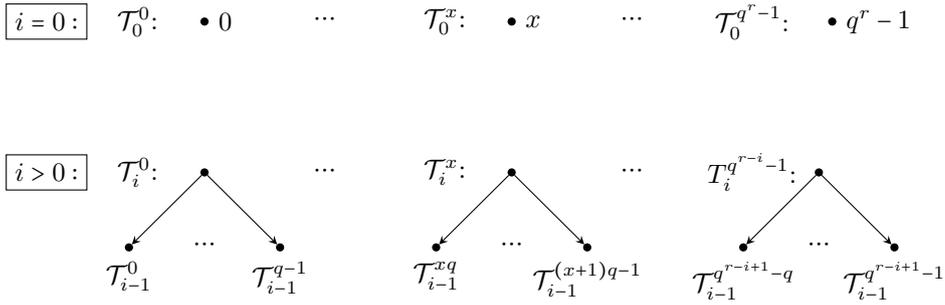
\begin{figure}[ht]
 \centering
\begin{tikzpicture}[>=stealth]
 
 %
 %
 
 %
 \node[draw,rectangle,thin] at (-3.2,0) {$i = 0:$};
 
 \node (t00) at (-2,0) {$\mcT_0^0$:};
 \node[right of=t00, node distance=25, draw,fill,circle, inner sep=0.8pt,
label=right:{$0$}] (treet00) {};
 
 \node[right of=treet00, node distance=45] (dotsib1) {$\cdots$};
  
  \node[right of=dotsib1, node distance=45] (t01) {$\mcT_0^x$:};
 \node[right of=t01, node distance=25, draw,fill,circle, inner sep=0.8pt,
label=right:{$x$}] (treet01) {};
 
 \node[right of=treet01, node distance=45] (dotsib2) {$\cdots$};
  
  \node[right of=dotsib2, node distance=45] (t0last) {$\mcT_0^{q^r-1}$:};
 \node[right of=t0last, node distance=30, draw,fill,circle, inner sep=0.8pt,
label=right:{$q^r - 1$}] (treet0last) {};

 %
 %
 
 %
 \node[draw,rectangle,thin] at (-3.2,-2) {$i > 0 :$};
 
 \node (ti0) at (-2,-2) {$\mcT_i^0$:};
 \node[right of=ti0, node distance=25, draw,fill,circle, inner sep=0.8pt]
(treeti0) {};
 \node[below left of=treeti0, node distance=40, draw, fill, circle,
inner sep=0.8pt,label=below:{$\mcT_{i-1}^{0}$}] (treeti0child1) {} ;
\node[below of=treeti0, node distance=28] (treeti0childdots) {$\cdots$} ;
  \node[below right of=treeti0, node distance=40, draw, fill, circle,
inner sep=0.8pt,label=below:{$\mcT_{i-1}^{q-1}$}] (treeti0childlast) {} ;
  
  \draw[->,thin] (treeti0) edge (treeti0child1);
  \draw[->,thin] (treeti0) edge (treeti0childlast);

 \node[right of=treeti0, node distance=45] (dotsis1) {$\cdots$};

 \node[right of=dotsis1, node distance=45] (tix) {$\mcT_i^x$:};
 \node[right of=tix, node distance=25, draw,fill,circle, inner sep=0.8pt]
(treetix) {};
 \node[below left of=treetix, node distance=40, draw, fill, circle,
inner sep=0.8pt,label=below:{$\mcT_{i-1}^{xq}$}] (treetixchild1) {} ;
\node[below of=treetix, node distance=28] (treetixchilddots) {$\cdots$} ;
  \node[below right of=treetix, node distance=40, draw, fill, circle,
inner sep=0.8pt,label=below:{$\mcT_{i-1}^{(x+1)q-1}$}] (treetixchildlast) {} ;
 
 \draw[->,thin] (treetix) edge (treetixchild1);
  \draw[->,thin] (treetix) edge (treetixchildlast);

 \node[right of=treetix, node distance=45] (dotsis2) {$\cdots$};
 
 \node[right of=dotsis2, node distance=45] (tilast) {$T_i^{q^{r-i}-1}$:};
 \node[right of=tilast, node distance=25, draw,fill,circle, inner sep=0.8pt]
(treetilast) {};
 \node[below left of=treetilast, node distance=40, draw, fill, circle,
inner sep=0.8pt,label=below:{$\mcT_{i-1}^{q^{r-i+1}-q}$}] (treetilastchild1)
{} ;
\node[below of=treetilast, node distance=28] (treetilastchilddots) {$\cdots$} ;
  \node[below right of=treetilast, node distance=40, draw, fill, circle,
inner sep=0.8pt,label=below:{$\mcT_{i-1}^{q^{r-i+1}-1}$}] (treetilastchildlast)
{} ;
  \draw[->,thin] (treetilast) edge (treetilastchild1);
  \draw[->,thin] (treetilast) edge (treetilastchildlast);
   
\end{tikzpicture}
\caption{Inductive definition of the trees $\mcT_i^x$}
\label{fig:definition:trees_q_sylow}
\end{figure}

We proceed with the inductive step $r \mapsto r+1$.
The set $\inseg q^{r+1}$ splits into $q$ blocks $\mcP_{r}^0, \dots,
\mcP_r^{q-1}$ each of size $q^{r}$.
The group $\Delta_r$ acts on $\mcP_{r}^0$ and point-wise fixes
the elements from the blocks $\mcP_{r}^x$ with $x \neq 0$.
Let $\gamma \in \Sym(\inseg n)$ for $n = q^{r+1}$ be the following 
permutation which shifts
the segments $\mcP_{r}^0, \dots, \mcP_r^{q-1}$ in a cycle of length $q$ by
composing the natural shifts on the sets of residues modulo $q^r$:
\[ \gamma = (0 \,  \cdots \, (q-1)q^r ) 
(1 \,  \cdots \, 1 + (q-1)q^r)  
\cdots
(q^r - 1 \,  \cdots \, q^r - 1 + (q-1)q^r).  
\]

Hence for all $a \in \inseg n $ we have $\gamma(a) = (a + q^r) \,\,\modulo 
q^{r+1}$.
We set $\Delta^{0}_r = \Delta_r$ and, more generally, $\Delta^{x}_r =
(\gamma^x) \Delta_r (\gamma^x)^{-1}$ for $x = 0, \dots, q-1$ to 
obtain $q$ copys of $\Delta_r$ which independently act on the segments
$\mcP_r^x$ for $0 \leq x \leq q-1$. 
Finally, we define $\Delta_{r+1}$ 
as the \emph{semi-direct product} of $(\Delta^{0}_r \times \cdots \times
\Delta^{q-1}_r)$ and the cyclic group $\gengroup{\gamma}$ of size $q$.
This means that the group elements of $\Delta_{r+1}$ are elements in the set
$(\Delta^{0}_r \times \cdots \times \Delta^{q-1}_r \times 
\gengroup{\gamma})$
and that the group operation is given by
\[ 
(\delta_1, \dots, \delta_{q-1}, \alpha) \cdot (\epsilon_1, \dots,
\epsilon_{q-1}, \beta) 
= (\delta_1 \cdot \alpha \epsilon_1 \alpha^{-1}, \dots, \delta_{q-1} \cdot
\alpha \epsilon_{q-1} \alpha^{-1}, \alpha\cdot \beta).
\]
Since $\card{\Delta_{r+1}} = \card{\Delta_r}^q \cdot q$ we conclude that
$\Delta_{r+1}$ indeed is a $q$-Sylow subgroup.

\medskip
From our construction it immediately follows that $\Delta_{r+1}$ satisfies the
properties stated in~(\ref{properties:sylow_tress:3}). To see that
$\Delta_{r+1}$ also satisfies the properties stated
in~(\ref{properties:sylow_tress:2}) we start by showing that, for $i \leq r$, 
$\Delta_{r+1}$ transitively acts on $\{\mcT_i^x : x \in \inseg {q^{r+1-i}} \}$.
If we split the set $\inseg {q^{r+1-i}}$  
into $q$ blocks $\mcP_{r-i}^0, \dots, \mcP_{r-i}^{q-1}$ of size $q^{r-i}$
then we know from the induction hypothesis that $\Delta_{r}^0$ transitively
acts on the set of trees $\{ \mcT_i^x : x \in \mcP_{r-i}^0 \} = \{
\mcT_i^x : x \in \inseg {q^{r-i}} \}$. 
Moreover, it is easy to verify that for all $x \in \inseg{q^{r+1-i}}$ we have
$\gamma(\mcT_i^x) = \mcT_i^z$ where $z = {x + q^{r-i} \mod q^{r+1-i}}$.
Hence $(\gamma^y) \{ \mcT_i^x : x \in \mcP_{r-i}^0 \} = \{ \mcT_i^x : x \in
\mcP_{r-i}^y \}$ for all $0 \leq y \leq q-1$ which means that
$\Delta_r^y$ transitively acts on $\{ \mcT_i^x : x \in \mcP_{r-i}^y \}$ and
thus~(\ref{properties:sylow_tress:2}) holds.

\bigskip
The crucial step is to understand the action of $\Delta_r$
on the sets $I_n = \inseg {n}^k$ and $J_n= \inseg{n}^\ell$ (for the case where 
$n = q^r$).
In fact, our next aim is to develop a complete invariant for 
the $\Delta_r$-orbits on these index sets. Recall that the sets 
of $\Delta_r$-orbits on $I_n$ and $J_n$ provide index sets for 
the succinct linear equation system 
$M_n^* \cdot \vec x = \onevec$.
To define this invariant, the main idea is to describe the position 
of a tuple $\ba \in I_n$ (or $\ba \in J_n$, respectively) in the tree 
$\mcT \defeq \mcT^0_r$.

Let us first define the \emph{signature} $\sgn(a,b)$ of a pair $(a,b) \in 
\inseg{n} \times \inseg{n}$ as the tuple $(i, z) \in \inseg{r+1} \times 
\inseg q$ such that the lowest common ancestor of $a, b$ in $\mcT$ is the 
root of a tree $\mcT^x_i$ and  such that $a$ is located in a subtree 
$\mcT_{i-1}^{xq + y_a}$ for $y_a \in \inseg q$ and $b$ is located in the 
subtree $\mcT_{i-1}^{xq + y_b}$ where $y_b = y_a + z \,\,\modulo q$. For the 
special case where $i=0$ we have $a = b$ and agree to set $z=0$. With this 
preparation we define the signature $\sgn(\ba)$ of a tuple $\ba = (a_1, 
\dots, a_\ell) \in J_n$ as the list $\sigma \in (\inseg{r+1} \times 
\inseg q)^{\ell (\ell -1) / 2}$ consisting of the individual signatures 
$\sgn(a_i,a_j)$ for all pairs $a_i, a_j$ with $1 \leq i < j \leq \ell$.
The signature of tuples in $I_n$ is defined analogously.

\begin{lemma}
\label{lemma:sgnsylow:part1}
 Let $\ba \in J_n$. Then $\sgn(\ba) = \sgn(\pi \ba)$ for all $\pi \in 
\Delta_r$.
\end{lemma}
\begin{proof}
 Immediately follows from the construction of $\Delta_r$ and the trees
$\mcT_i^x$.
\end{proof}

\begin{lemma}
Let $\ba, \bb \in J_n$. If $\sgn(\ba) = \sgn(\bb)$, then $\bb \in
\Delta_r(\ba)$.
\end{lemma}
\begin{proof}
 We proceed by induction on the maximal position $0 \leq i \leq \ell$ such that
$a_j = b_j$ for all $j=1, \dots, i$. The case $i = \ell$ is clear, so assume
that $i < \ell$.
Let $\ba = (a_1, \dots, a_i, a_{i+1}, \dots, a_\ell)$ and $\bb = (a_1, \dots,
a_i, b_{i+1}, \dots, b_\ell)$. We show that there exists a permutation $\delta
\in \Delta_r$ which pointwise fixes $a_1, \dots, a_i$ and such that
$\delta(a_{i+1}) = b_{i+1}$. Then the claim follows from
Lemma~\ref{lemma:sgnsylow:part1} together with the induction hypothesis.
For $i = 0$ this is easy, because $\Delta_r$ acts transitively on $\inseg
n$. If $i > 0$ we choose $a_w \in \{ a_1, \dots, a_i \}$ such that
$\sgn(a_w, a_{i+1}) = (c,d)$ and such that $c$ is minimal with this 
property.
Obviously we
have $c > 0$. By the choice of $a_w$ the lowest common ancestor of
$a_w$ and $a_{i+1}$ is the root of a tree $\mcT_c^x$. Moreover, $a_w$ is
located in a subtree $\mcT_{c-1}^{xq +y}$ for some $0 \leq y \leq q-1$ and
$a_{i+1}$ is located in the subtree $\mcT_{c-1}^{xq +z}$ where $z = y + d
\mod q$. Since $\sgn(\ba) = \sgn(\bb)$ also $b_{i+1}$ occurs as the label of a
leave in the subtree $\mcT_{c-1}^{xq +z}$.
By the minimality assumption on $c$ we know that non of the elements 
$\{ a_1, \dots, a_i \}$ occurs in the tree $\mcT_{c-1}^{xq +z}$.
Hence, by the
properties of the group $\Delta_r$ stated in~(\ref{properties:sylow_tress:3})
we can find an element $\delta \in \Delta_r$ which point-wise fixes all elements
outside the block $\mcP_{c-1}^{xq +z}$ (in particular, the elements $a_1, \dots,
a_i$) and which moves $a_{i+1}$ to $b_{i+1}$. 
\end{proof}

Following our definition from above, the signature $\sgn(\ba)$ of an element 
$\ba \in J_n$ is a tuple of length $\ell (\ell -1) / 2$ whose entries are pairs 
$(i, z) \in \inseg{r+1} \times  \inseg q$. We denote the set of all possible 
sequences of this form by $S^\ell_n = (\inseg{r+1} \times \inseg q)^{\ell (\ell 
-1) / 2}$. Of course, not every tuple in $\sigma \in S_n^\ell$ can be
realised as the signature $\sgn(\ba)=\sigma$ of an element $\ba \in J_n$. 
Similarly, we define the set $S^k_n = (\inseg{r+1} \times \inseg q)^{k (k 
-1) / 2}$ to capture all possible signatures of elements in $I_n$.

\medskip
Since the coefficient matrix $M_n^*$ of the equivalent linear 
equation system $M_n^* \cdot \vec x = \onevec$ can be defined 
as a matrix whose index sets are the collections of $\Delta_r$-orbits on 
$I_n$ and $J_n$, we can use the notion of signatures to describe 
$M_n^*$ as an 
$(S_n^k  \times S_n^\ell)$-matrix. This fits with our 
proof plan as the index sets $S_n^k$ and $S_n^\ell$ of the 
matrix $M_n^*$ are much smaller than the  index sets $I_n$ and $J_n$ of the 
coefficient matrix $M_n$ of the original linear equation system.
However, it still might be the case that the succinctness of the 
matrix $M_n^*$ does not help, because it is not possible to obtain its 
entries within $\FORp$.

We show that this is not the case. More precisely we show
that we can define the matrix $M_n^*$ in $\FOC$ in a 
structure of size $r$ (where we assume that $r \geq q$). 
Therefore, the main technical 
step is to show that $\FOC$ can count (modulo $p$) the number of 
realisations of a potential signature $\sigma \in S_n^k$.

\medskip
First of all, we need some further notation. A \emph{complete equality 
type in $k+\ell$ variables}  is a consistent set $\tau(x_1, \dots, x_k, 
x_{k+1}, \dots, x_{k+\ell})$ of literals $x_i = x_j, x_i \neq x_j$
which contains for every pair $i < j$ either the atom $x_i = x_j$ or the literal
$x_i \neq x_j$. Note that each quantifier-free formula $\alpha \in 
\FO(\emptyset)$ can be 
expressed as a Boolean combination of complete equality types.

In the following main technical lemma we show that in the structure $\mfA = 
(\inseg{ r})$ we can count (modulo $p$) the number of 
realisations of a (potential) signature $\sigma \in S_n^\ell$ in a 
subtree $\mcT_i^x$ in $\FOC$. More generally, this is possible  if 
we additionally 
fix some entries of the tuples which should realise $\sigma$ in 
$\mcT_i^x$.
Here we need another prerequisite: as we want to work with elements 
from the set $\inseg n = \inseg {q^r}$ in a structure of size $r$ we have to 
agree on some sort of succinct representation. 
Of course the natural choice here is 
to represent numbers $x \in \inseg n$ in the structure $\mfA$ via their 
$q$-adic encoding: a binary relation $R \subseteq 
\inseg{r}^2$ which corresponds to a function $R: \inseg{r} \to \inseg{q}$ 
represents the number $x(R) \in \inseg{n} = \sum_{i=0}^{r-1} R(i) \cdot 
q^i$. Note that this encoding requires a linear order on the set $\inseg r$ 
(which is \emph{not} the case for the 
structure $\mfA$). However, as we are working with $\FOC$ we can just use the 
number sort on which a linear order is available.
Hence in the following, whenever we specify $\FOC$-formulas or $\FOC$-terms 
with free variables or with free relation symbols which should represent 
numbers, then we implicitly assume that these variables are \emph{numeric} 
variables and that the relation symbols are evaluated over the number 
sort. The same holds for signatures $\sigma \in S_n^\ell$ which we specify 
in $\FOC$-formulas by a list of pairs $(h_i, d_i)$ of \emph{numeric} 
variables  of length $\ell \choose 2$.

Before we state our main technical lemma it is helpful to recall that our 
inductive construction of the trees $\mcT_i^x$ fits very well with the 
$q$-adic encoding 
of numbers $x \in \inseg n$.
Again, let $x \in \inseg{n}$ be given by its $q$-adic encoding as 
$x = (x_0, \dots, x_{r-1}) \in \inseg{q}^r$, i.e.\ $ x = \sum_{i = 
0}^{r-1} x_i \cdot q^i$. Then the $i$-th node on the unique path from the root 
in the tree $\mcT = \mcT_r^0$  to the leave $\mcT_0^x$ is the root of the tree 
$\mcT_{r-i}^{y}$ where $y = x_{r-i} + x_{r-i+1} q + \cdots + x_{r-1} q^{i-1}$. 
In other words, the $q$-adic encoding of $x$ precisely describes the 
path in the tree $\mcT$ from the root to the leave labelled with $x$ where 
at level $(r-i)$ the $i$ last entries $x_{r-i}, \dots, x_{r-1}$  in 
the $q$-adic encoding of $x$ are determined (i.e.\ $x$ is a member of the 
block $\mcP_{r-i}^y$).

\begin{lemma}\label{lemma:counting:realisations:sgn}
  For all $\ell \geq 1$ and $0 \leq s \leq \ell$ there exist
 \begin{enumerate}[(a)]
  \item a term $\Theta(i,h_1,d_1, \dots, h_t, d_t) \in \FOC(\{ 
R_x, R_1, \dots, R_s \})$, and
  \item formulas $\phi_e(y,z,i,h_1,d_1, \dots, h_t,d_t) \in 
\FOC(\{ R_x, R_1, \dots, R_s \}) $ for $e = s+1, \dots, \ell$,
 \end{enumerate}
where $t = {\ell \choose 2}$, such that for all $r \geq q$, all $i \leq 
r$, all 
$\sigma=((h_1, d_1), \dots, (h_t, d_t)) \in S_n^\ell$ 
where $n = q^r$, all $x \in \inseg{q^{r-i}}$ and all $a_1, \dots, a_s 
\in \mcP_i^x$ the following holds: let $\mfA = (\inseg{r})$ and let $R_x, 
R_1, \dots, R_s$ be numerical relations such that $R_x$ represents the 
($q$-adic 
encoding of the) element $x \in \inseg{q^{r-i}}$ and such that each $R_i$ 
represents 
the  ($q$-adic encoding of) the element $a_i$. Then we have that
 \begin{enumerate}[(i)]
  \item the value $\Theta^\mfA(q,i,h_1,d_1, \dots, h_t, d_t)$ of the 
term $\Theta$ in $\mfA$ is $\card{Z} \,\,\modulo p$ where 
\[Z = \{ (a_{s+1}, \dots, a_\ell) \in (\mcP_i^x)^{ \ell -s } : \sgn(a_1, 
\dots, a_s, a_{s+1}, \dots, a_\ell) = 
\sigma \}. \]
 \item if $Z \neq \emptyset$, then the formulas $(\phi_e)_{s < e \leq 
\ell}$ define the $q$-adic representation of witnessing elements $a_{s+1}, 
\dots, a_\ell \in \mcP_i^x$, i.e.\ such that $(a_{s+1}, \dots, a_\ell) \in 
Z$.
\end{enumerate}

\end{lemma}

\begin{proof}
 First of all, by our previous observations it is easy to see that the 
condition $a_j \in \mcP_i^x$ for $j = 1, \dots, s$ can be defined in 
$\FOC$. More generally, we can use the $q$-adic encoding of the elements 
$a_j$ to determine $\sgn(a_1, \dots, a_s)$ in $\FOC$. Hence, for the 
remainder of the proof we assume that $\sgn(a_1, \dots, a_s)$ is consistent 
with $\sigma$ and that $a_j \in \mcP_i^x$ for $j = 1, \dots, s$.

We proceed by induction on $\ell$. For $\ell = 1$ it suffices to 
show that $\FOC$ can compute $(n \,\,\modulo p)$ where $n =q^r$ in the 
structure $\mfA$. To see this, recall that $p$ and $q$ are co-prime and 
thus we can use Lagrange's theorem to conclude that $q^r \equiv q^{r'} 
\,\,\modulo p$ if $r' \equiv r \,\,\modulo (p-1)$. 
Since $p$ is a constant, the claim follows.

Let $\ell \geq 2$. We distinguish between the following two cases. If $s = 0$, 
then we can partition the set of realisations $\ba$ of $\sigma$ according 
to first entry $a_1$ into $\card{\mcP_i^x}$ parts of 
equal size.
It suffices to determine the size of each of these blocks, since we can
determine $\card{\mcP_i^x} \,\,\modulo p$
in $\FOC$ similarly as above.

Without loss of generality let us assume that $a_1 = x \cdot q^i$. Since 
we 
have given the $q$-adic encoding of $x$ it is easy to see that we can 
define the $q$-adic encoding of $xq^i$ in \FOC. This gives us the formula 
$\phi_1$.
Next, we partition the set of indices $\{ 2, \dots, \ell \}$ into classes 
according to the
equivalence relation
$j_1 \approx j_2$ if $\sigma(1,{j_1}) = \sigma(1,{j_2})$. Let the
resulting
classes be $Y_1, \dots, Y_v$ and let $\sigma(1,y) = (h_w, d_w)$ for all
$y \in Y_w$ and $w = 1, \dots, v$. 

We observe that there exists a tuple $\ba$ with $a_1 = x \cdot q^i$ which
realises $\sigma$ in the tree $\mcT_i^x$ (that is $Z \neq \emptyset$)
if, and only if, the following conditions are satisfied:

\begin{itemize}
 \item for all $w = 1, \dots, v$ we have $h_w \leq i$, and
 \item for every $Y_w = \{ y^w_1, \dots, y^w_{\ell_w} \}$ there is a 
tuple $\ba^w$ of length $\ell_w$ which realises $\sigma$ (restricted to the
indices from $Y_w$) in the subtree $\mcT_{h_w-1}^{xq^{i-h_w+1}+d_w}$, and
 \item for all pairs $y_1 \in Y_{w_1}$ and $y_2 \in Y_{w_2}$ with $w_1 
\neq w_2$ we
have that 
  \[\sigma({y_1},{y_2}) = \begin{cases}
        (h_{w_1},d_{w_2} - d_{w_1} \,\,\modulo q), & \text{if } h_{w_1} = 
h_{w_2} 
\\
        (h_{w_2},d_{w_2}) & \text{if } h_{w_1} < h_{w_2} \\
        (h_{w_1},d_{w_1}) & \text{if } h_{w_2} < h_{w_1}.
                  \end{cases} \]
\end{itemize}

Since $\ell$ is a constant, the number of possible partitions of $\{2,
\dots, \ell \}$ is bounded by a constant as well. It is easy to see that for
every possible such partition we can check the first and third
condition in $\FOC$. 
To verify the second condition in $\FOC$ we use the induction 
hypothesis. 
There are two aspects which have to be discussed with more precision. First of
all, we have to handle one particular case separately: indeed, if $h_w = 1$
for all $w = 1, \dots, v$, then we cannot use the induction
hypothesis since all elements (including $a_1$) have to be chosen in the 
same subtree of height one.
However, in this case there is only one realisation (if the third condition is
satisfied) so this does not cause any problems.
The other difficulty is that we have to define the $q$-adic encoding of the
value $z_w={xq^{i-h_w+1}+d_w}$ in $\FOC$. 
We already noted before 
that the $q$-adic representation of $xq^{i-h_w+1}$ can be defined in $\FOC$ and
since $0 \leq d_w < q$ we can also define the $q$-adic encoding of $z$ in
$\FOC$.

In fact, the induction hypothesis also provides us with a term which counts
modulo $p$ the number of possible realisations of $\sigma$ in the subtrees 
$\mcT_{h_w-1}^{z_w}$ restricted to the indices in $Y_w$ together
with formulas $\phi_e$ which define witnessing elements. 
Finally, since the overall number of possible realisations of $\sigma$ in
$\mcT_i^x$ is the product of the realisations restricted to the components
$Y_w$, the claim follows for the case where $s = 0$. 

For the general case let $\ell \geq s > 0$ and let $a_1, \dots, a_s \in
\mcP_i^x$ be the components of the tuple $\ba$ that are already fixed. 
Recall that we can assume without loss of generality that 
$\sgn(a_1, \dots, a_s)$ is consistent with $\sigma$ and that
all elements $a_1, \dots, a_s$ are located in the subtree $\mcT_i^x$.
Since we have fixed the element $a_1$, we can proceed as above except for two
small changes. First of all, when applying the induction hypothesis we have 
to respect the remaining fixed elements $a_2, \dots, a_s$. Moreover, when we
form the partitions of $\{2, \dots, \ell \}$ into parts $Y_1, \dots, Y_v$ as
above then we have to adapt the position of elements corresponding to the index
set $Y_w$ since the element $a_1$ is not necessarily contained in the tree
$\mcT_{h_w+1}^{xq^{i-h_w+1}}$. However, since we have given the $q$-adic
representation of $a_1$ we can define in $\FOC$ the element $0 \leq d_a < q$
such that $a_1$ is located in the subtree $\mcT_{h_w+1}^{xq^{i-h_w+1}+d_a}$. 
The remaining steps can be performed as above. This finishes our proof.
\end{proof}

\begin{lemma}\label{lemma:entrymn:comp:type}
Let
$\tau(x_1, \dots, x_k, y_1, \dots, y_\ell)\in \FO(\emptyset)$ be a complete 
equality type (in $k + \ell$ variables).
Then there is an $\FOC$-term $\Theta_\tau(\bz_x,\bz_y)$ such that for all 
$r \geq q$, all $\sigma_{\ba} \in S_n^k$ and $\sigma_{\bb} \in 
S_n^\ell$, 
where $n = q ^r$, the value
$\Theta^\mfA(\sigma_{\ba}, \sigma_{\bb})$ of $\Theta$ in $\mfA =
(\inseg r)$ is 
\[\Theta^\mfA(q, \sigma_{\ba},
\sigma_{\bb})= \card{ \{ \bb \in J_n : \sgn(\bb) = \sigma_{\bb}, (\inseg 
n) \models
\tau(\ba,\bb) \}} \,\,\modulo p\]
for some (or, equivalently, all) $\ba \in I_n$ with $\sgn(\ba) = \sigma_{\ba}$.
\end{lemma}
\begin{proof}
 By Lemma~\ref{lemma:counting:realisations:sgn} we can first check in
$\FOC$ that $\sigma_{\ba}$ and $\sigma_{\bb}$ can be realised (otherwise the
answer is trivial). Moreover, if $\tau$ (restricted to $x_1, \dots, x_k$) is not
consistent with $\sigma_{\ba}$ or if $\tau$ (restricted to $y_1, \dots, y_\ell$)
contradicts $\sigma_{\bb}$, then the answer is trivial as well.

In all other cases, Lemma~\ref{lemma:counting:realisations:sgn} provides 
$\FOC$-formulas which define in the structure $\mfA$ the $q$-adic encoding 
of elements $a_1, \dots, a_k \in \inseg{n}$ such that $\sgn(\ba) = 
\sigma_{\ba}$. 
Moreover, if $\tau$ contains a literal $x_i = y_j$, then we can
fix the entry $b_j$ as well. Hence, let us assume without loss of 
generality that $\tau$ contains the literals $x_i \neq y_j$ for all 
$1 \leq i \leq k$ and $1 \leq j \leq \ell$.

For $Y \subseteq \{ 1, \dots, \ell \}$ and a partial assignment $\epsilon: \{1,
\dots, \ell \} \to \{ a_1, \dots, a_k \}$ with $\dom(\epsilon) \cap Y =
\emptyset$ we define the set 
\[ B_Y^\epsilon = \{ \bb \in J_n : \sgn(\bb) = \sigma_{\bb}, \text{ for } i \in
\dom(\epsilon): b_i = \epsilon(i), \text{ for } i \in Y: b_i \neq a_1,
\dots, a_k \}.  \] 

With this notation our overall aim is to determine $(\card{B_{Y }^\emptyset}
\,\,\modulo p)$ for $Y = \inseg \ell$ in $\FOC$. The first observation is
that by Lemma~\ref{lemma:counting:realisations:sgn} we can determine
$(\card{B_{\emptyset}^\epsilon} \,\,\modulo p)$ for all partial
assignments $\epsilon$ in $\FOC$.
The second observation is that we can construct the values
$(\card{B_{Y}^\epsilon} \,\,\modulo p)$ by induction on $\card{Y}$ as follows. 
For
$Y \subseteq \{1, \dots, \ell \}$ and a partial assignment $\epsilon$ (with
$\dom(\epsilon) \cap Y = \emptyset$) we have for all $j \in Y$ that
\[
\card{B_Y^\epsilon} = \card{B^\epsilon_{Y\setminus \{ j \}}} - 
\sum_{a \in \{a_1, \dots, a_k \}} 
\card{ B^{\epsilon \, \cup \{ j \mapsto a \} }_{Y\setminus \{ j \}} } .
\]
In this way we recursively obtain the value 
$(\card{B_{Y }^\emptyset} \,\,\modulo p)$ for $Y = \inseg \ell$. 
Since $\ell$ is a 
constant the recursion depth is bounded by a constant as well and the 
procedure can be formalised in $\FOC$. 
\end{proof}

\begin{lemma}\label{lemma:mnstar:definable}
There exists an $\FOC$-term $\Theta(\bmu, \bnu)$ which defines for 
all $r \geq q $ in the structure $\mfA = (\inseg{r})$ the matrix $M_n^*$ 
where $n = q^r$.
\end{lemma}

\begin{proof}
 Recall that we can view $M_n^*$ as an $(S_n^k \times 
S_n^\ell)$-matrix over $\field F_p$. To represent the index sets $S_n^k$ 
and $S_n^\ell$ we let $\bmu$ and $\bnu$ be tuples of numeric variables of 
lengths $|\bmu| = {k \choose 2}$ and $|\bnu|  = {\ell \choose 2}$, 
respectively.

Note that the entry $M_n^*(\sigma_{\ba}, \sigma_{\bb})$ of $M_n^*$
for $\sigma_{\ba} \in S_n^k$ and $\sigma_{\bb} \in S_n^\ell$ is given as 
\[ M_n^*(\sigma_{\ba},\sigma_{\bb}) = \card { \{ \bb \in J_n : \sgn(\bb) 
= \sigma_{\bb}, M_n(\ba,\bb) = 1 } \mod p, \]
for some (or, equivalently, all) $\ba \in I_n$ with $\sgn(\ba) = 
\sigma_{\ba}$. The entry $M_n(\ba,\bb)$, in turn, is 
determined by the quantifier-free formula $\alpha(\bx_1, \bx_2) \in 
\FO(\emptyset)$. With this preparation, 
Lemma~\ref{lemma:entrymn:comp:type} already shows that we can determine 
the value $M_n^*(\sigma_{\ba},\sigma_{\bb})$ for the case where $\alpha$ is 
a complete equality type. For the general case we just write $\alpha$ as 
the union of complete equality types and combine the constant number of 
intermediate results.
\end{proof}

\begin{definition}
 Let $\mcK \subseteq \Str(\emptyset)$ be a class of sets. The 
\emph{$q$-power $\mcK^q \subseteq \Str(\emptyset)$ of $\mcK$} 
consists of all sets $\mfA = (\inseg {q^r})$ such that $\mfB = (\inseg r) 
\in \mcK$.
\end{definition}

\begin{theorem}
\label{thm:foslv:qred}
 Let $\mcK \subseteq \Str(\emptyset)$ be a class of sets. If $\mcK^q$ is 
definable in $\FOSp$, then 
$\mcK$ is definable in $\FORp$.
\end{theorem}
\begin{proof}
 If $\mcK^q$ is definable in $\FOSp$, then by 
Theorem~\ref{thm:normalform:fos} we can also find a formula $\phi  = 
(\slvp \, \bx_1,\bx_2) \alpha(\bx_1,\bx_2) \in 
\FOSp$ 
that defines $\mcK^q$ such that 
$\alpha$ is quantifier-free. 

By using the above construction and Lemma~\ref{lemma:mnstar:definable}, 
we conclude that the linear equation system $M_n \cdot \vec x = \onevec$ 
defined by $\alpha$ in an input structure $\mfA = (\inseg n)$ of size 
$n = q^r$ can be transformed into the equivalent 
system $M_n^* \cdot \vec x = \onevec$ which is $\FOC$-definable in $\mfB = 
( \inseg r)$.
Let $\phi^* \in \FORp$ be a formula which expresses the solvability of 
the linear system $M_n^* \cdot \vec x = \onevec$ in a structure $\mfB = 
(\inseg r)$.

Then $\mfB \models \phi^*$ if, and only if, $\mfA \models \phi$ since the 
linear equation systems $M_n \cdot \vec x = \onevec$ 
and $M_n^* \cdot \vec x = \onevec$ are equivalent.
\end{proof}

\begin{theorem}
 For all primes $p$ we have 
 $\FOSp < \FORp$ (even over $\Str(\emptyset)$).
\end{theorem}
\begin{proof}
 Otherwise we had $\FOSp = \FORp$. As above we fix some prime $q \neq p$. 
Let $\mcK \subseteq \Str(\emptyset)$ be a class of sets such that $\mcK 
\nin \FORp$, but such that $(\mcK^q)^q \in \FORp$. Such a class $\mcK$ is 
well-known to exist (just combine the fact that, over sets, we have 
$\LOGSPACE \leq \FORp \leq \PTIME$ and the space-hierarchy 
theorem, see e.g.~\cite{Pa95}). Since $\FOSp = \FORp$ we had $(\mcK^q)^q 
\in 
\FOSp$ and by Theorem~\ref{thm:foslv:qred} this means that $\mcK^q \in 
\FORp$. Again, since $\FORp = \FOSp$, we had $\mcK^q \in \FOSp$. A second 
application of Theorem~\ref{thm:foslv:qred} yields $\mcK \in \FORp$ 
which contradicts our assumptions.
\end{proof}

Finally we remark that the proof also works for the extension of 
fixed-point logic by solvability quantifiers but in the absence of counting.
The simple reason is that fixed-point operators do not increase the expressive 
power of first-order logic over the empty signature since all definable 
relations are composed from a constant-sized set of basic building blocks.

\section{Discussion}

We showed that the expressive power of rank operators over 
different prime fields is incomparable and we inferred that the version of rank 
logic \FPR with a distinct rank operator $\rkp$ for every prime $p \in \bbP$ 
fails to capture polynomial time.
In particular our proof shows that $\FPR$ cannot express the uniform version of 
the matrix rank problem where the prime $p$ is part of the input.
Moreover, we separated rank operators and solvability quantifiers in the 
absence of counting.

Of course, an immediate question is whether the extension $\FPRvar$ of 
$\FPC$ by the uniform rank operator $\rk$ suffices to capture polynomial time.
We do not believe that this is the case.
A natural candidate to separate $\FPRvar$ from $\PTIME$ is the solvability 
problem for linear equation systems over finite rings
rather than fields~\cite{DaGrHoKoPa13}.
While linear equations systems can be efficiently solved also over rings,
there is no notion of matrix rank that seems to be helpful for this purpose.
In particular, it is open, whether $\FPRvar$ can
define the isomorphism problem for CFI-structures 
generalised to $\Z_4$.
A negative answer to this last question would provide a class 
of structures on which $\FPRvar$ is strictly weaker than
Choiceless Polynomial Time (which captures $\PTIME$ on this 
class~\cite{AGGP14}).

Another question concerns the relationship between solvability 
logic $\FPS$ and rank logic $\FPRvar$. Our proof of 
Lemma~\ref{lemma:rank:to:solve} shows that on every class of structures of 
bounded colour class size the two logics have the same expressive power.
However, over general structures this reduction fails.
We only know, by our results from Section~\ref{sec:slv:rk}, that
a simulation of rank operators by solvability quantifiers would require
counting.

Finally, we think it is worth to explore the connections between our approach 
and the game-theoretic approach proposed by Dawar and Holm in~\cite{DaHo12} to 
see to what extent our methods can be combined. 
For example, what kind of properties does a variant of their partition games 
have for  infinitary logics with solvability quantifiers?

\bibliography{fps-fpr}

\end{document}